%% file: cen.tex
\documentclass[usenatbib]{mn2e}

\pdfoutput=1

\usepackage{fixltx2e}
\usepackage{mathptmx}
\usepackage{url}
\usepackage[pdftex]{graphicx}

\include{defn}

\begin{document}

\title[Cool gas in the Centaurus cluster]
{Cool X-ray emitting gas in the core of the Centaurus cluster of
  galaxies}
\author[J.S. Sanders et al.]
{\parbox[]{6.in}{{J.~S. Sanders$^1$\thanks{E-mail: jss@ast.cam.ac.uk},
      A.~C. Fabian$^1$, S.~W. Allen$^2$, R.~G. Morris$^2$, J. Graham$^1$\\ and
      R.~M. Johnstone$^1$}\\
    \footnotesize
    $^1$ Institute of Astronomy, Madingley Road, Cambridge CB3 0HA\\
    $^2$ Kavli Institute for Particle Astrophysics and Cosmology,
    Stanford University, 382 Via Pueblo Mall, Stanford, CA 94305-4060,
    USA
  }
}
\maketitle

\begin{abstract}
  We use a deep \emph{XMM-Newton} Reflection Grating Spectrometer
  observation to examine the X-ray emission from the core of the
  Centaurus cluster of galaxies. We clearly detect Fe~\textsc{xvii}
  emission at four separate wavelengths, indicating the presence of
  cool X-ray emitting gas in the core of the cluster. Fe ions from
  Fe~\textsc{xvii} to \textsc{xxiv} are observed. The ratio of the
  Fe~\textsc{xvii} 17.1{\AA} lines to 15.0{\AA} line and limits on
  O~\textsc{vii} emission indicate a lowest detected temperature in
  the emitting region of 0.3 to 0.45~keV (3.5 to $5.2 \times
  10^6$~K). The cluster also exhibits strong N~\textsc{vii} emission,
  making it apparent that the N abundance is supersolar in its very
  central regions. Comparison of the strength of the Fe~\textsc{xvii}
  lines with a Solar metallicity cooling flow model in the inner
  17~kpc radius gives mass deposition rates in the absence of heating
  of $1.6-3 \Msunpyr$. Spectral fitting implies an upper limit of
  $0.8\Msunpyr$ below 0.4~keV, $4 \Msunpyr$ below 0.8~keV and
  $8\Msunpyr$ below 1.6~keV.  The cluster contains X-ray emitting gas
  over at least the range of 0.35 to 3.7~keV, a factor of more than 10
  in temperature. We find that the best fitting metallicity of the
  cooler components is smaller than the hotter ones, confirming that
  the apparent metallicity does decline within the inner 1~arcmin
  radius.
\end{abstract}


\section{Introduction}
In the centres of many clusters of galaxies the mean radiative cooling
time of the intracluster medium (ICM) is short, typically less than
1~Gyr (e.g. \citealt{Voigt04}). This is also where the ICM is
significantly cooler than the outskirts. It is therefore expected that
a cooling flow \citep{Fabian94} should be formed, where material cools
out of the ICM.

High spectral resolution X-ray studies of nearby clusters of galaxies
using the Reflection Grating Spectrometer (RGS) instruments on
\emph{XMM-Newton} has revealed a lack of cool X-ray emitting gas in
these objects \citep{Tamura01b, Tamura01a, Peterson01, Kaastra01,
  Sakelliou02, Peterson03}.  There are emission lines observed from
gas down to a half or a third of the outer temperature, but with very
little gas at cooler temperatures (see \citealt{PetersonFabian06} for
a review). The main spectral indicator of the lack of cool gas is the
weakness of Fe~\textsc{xvii} lines.

A cluster where there has been tentative evidence for the existence of
cool gas is Abell~2597 \citep{MorrisFabian05}, which shows possible
Fe~\textsc{xvii} emission in its RGS spectrum. NGC~4636 emits strong
Fe~\textsc{xvii} lines, but these are consistent with a small range of
temperature as the group is cool \citep{Xu02}. NGC~5044 is another
cool system showing strong Fe~\textsc{xvii} and hinting at
O~\textsc{vii} emission \citep{Tamura03}.

The Centaurus cluster, Abell~3526, is a nearby bright cluster of
galaxies. It lies at a redshift of 0.0104
\citep{LuceyCurrieDickens86a}, and a 2-10 keV X-ray luminosity of $2.9
\times 10^{43} \ergps$ \citep{Edge90}. It lies at a relatively low
Galactic latitude of $21.6^\circ$ with a Hydrogen column density
towards the cluster of $\sim 8.6 \times 10^{20} \psqcm$ (derived from
the H~\textsc{i} maps of \citealt{Kalberla05}).

A cool X-ray emitting gas component has been seen in this object by
low spectral resolution CCD quality observations. This component was
first seen by \emph{ASCA} \citep{Fukazawa94} and \emph{ROSAT}
\citep{AllenFabian94}, who found cool gas with a temperature of $\sim
1$~keV in addition to a hot $\sim 4$~keV component.  \cite{Ikebe99}
put forward a two-phase model fitting \emph{ASCA} and \emph{ROSAT}
data, claiming 1.4 and 4.4~keV components over the inner 3~arcmin.  A
colour analysis of \emph{ROSAT} data \citep{Sanders00} gave a mean
temperature of 1.7~keV in the central 90~arcsec. \cite{Allen01} found
an improved fit to their \emph{ASCA} spectra with two or more
temperatures, with temperatures at 3.2 and 1.3~keV. There was also
evidence for a hot or powerlaw component (seen previously by
\citealt{Allen00} and \citealt{DiMatteo00}). \emph{BeppoSAX}
observations \citep{Molendi02} found a peak temperature of 4~keV,
dropping to below 2~keV in the core, and evidence for a hard component
in the PDS instrument.

The excellent spatial resolution of \emph{Chandra} provided a more
complete picture of the temperature of the gas in the central regions
\citep{SandersCent02}. This observation showed a plume-like feature in
the core of the cluster, emitting soft X-rays. Two-temperature
component fits to projected spectra extracted from the plume were
preferred to single-temperature fits, with temperatures of 0.7 and
1.5~keV. Outside this region the temperature increases quickly to the
east to around 3.7~keV. To the west, there is a plateau of cooler gas
at around 2.5~keV, before the temperature rises at a radius of around
190~arcsec. Deeper \emph{Chandra} observations \citep{Fabian05}
confirmed this picture, highlighting the clear east-west asymmetry of
the temperature distribution. \emph{XMM-Newton} EPIC data confirmed
the $\sim 0.7\keV$ component in the core \citep{SandersEnrich06}, also
finding the temperature of the ICM to drop to around 3.4~keV beyond
120~kpc radius. The \emph{Chandra} observations also clearly showed
that the metallicity of the ICM is inversely correlated with its
temperature \citep{SandersCent02,Fabian05,SandersEnrich06}.

An interesting feature of this cluster is that the metallicity of the
gas towards the centre is significantly supersolar. The Fe metallicity
peaks between 1.5 and 2\Zsun \citep{Fabian05,SandersEnrich06}. Si and
S peak around 2\Zsun, and Ni peaks around 4\Zsun. In addition the
metallicity of the gas appears to decline in the very central regions
\citep{SandersCent02}.

The multiple detections of cool X-ray emitting gas in the core of
Centaurus provides an excellent opportunity to test the picture that
there is only a range in 2--3 in X-ray gas temperature in clusters of
galaxies. We therefore undertook a deep \emph{XMM-Newton} RGS
observation of Centaurus, which, when combined with the existing
observation, gives a total exposure of around 160~ks.

In this paper we assume $H_0 = 70 \kmpspMpc$, which gives an angular
scale of 213~pc per arcsec for Centaurus. We assume the Solar
metallicities of \cite{AndersGrevesse89}. Error bars are quoted as
1$\sigma$ and limits as 2$\sigma$.

\section{Data preparation}
\begin{figure*}
  \includegraphics[width=\textwidth]{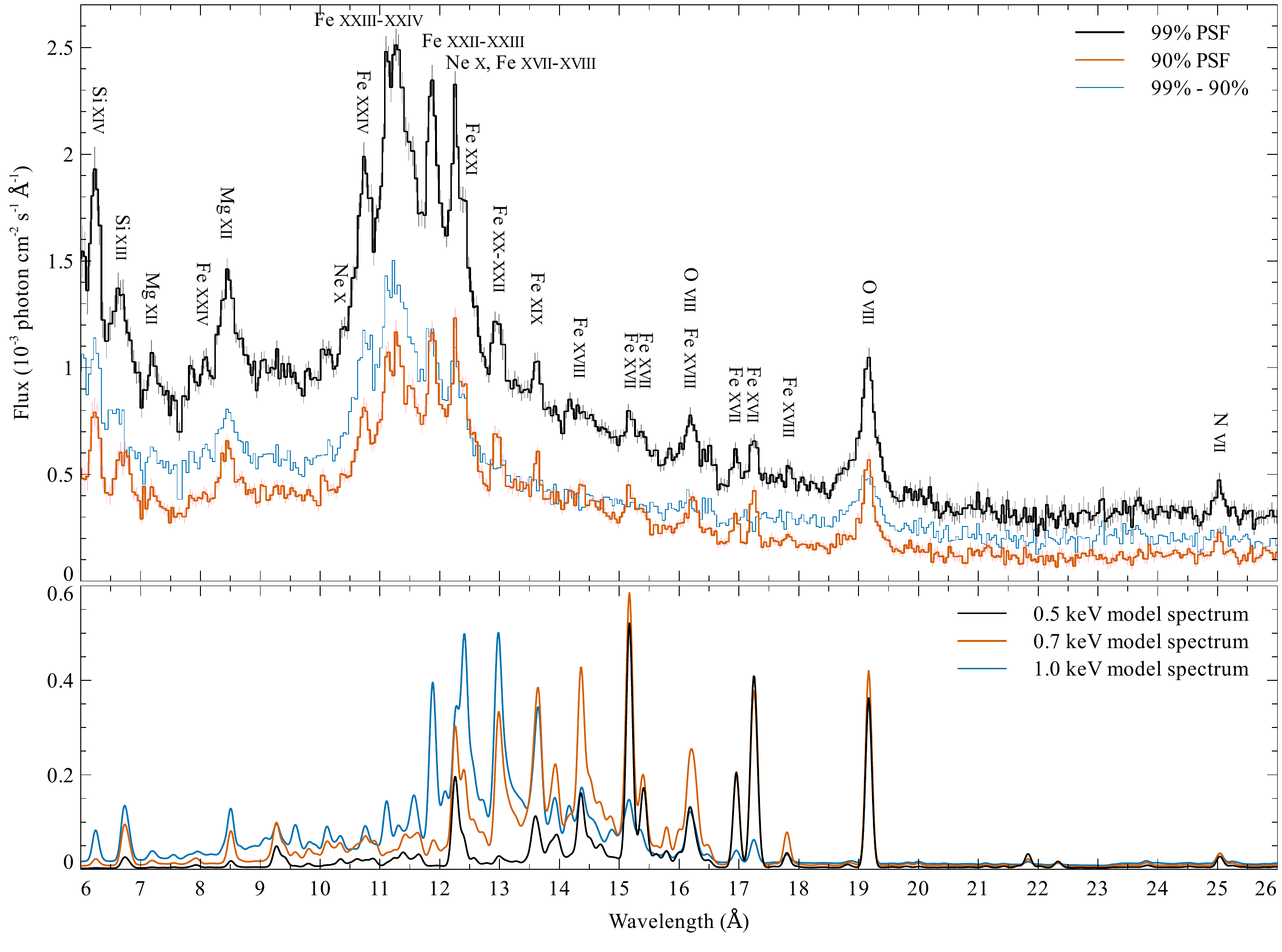}
  \caption{Top panel: fluxed and combined 1st and 2nd order RGS1 and
    RGS2 data from the two observations. Shown are the spectrum using
    95 and 99 per~cent of the RGS PSF. Also plotted is the difference
    between the two spectra, showing the emission from outside of the
    very central region. The spectra have been binned by a factor of 4
    to have 0.04{\AA} spectral bins. Bottom panel: smoothed
    \textsc{apec} Solar model spectra for gas at 0.5, 0.7 and 1~keV
    with arbitrary normalisation.}
  \label{fig:spectrum}
\end{figure*}

We processed the two source datasets listed in
Table~\ref{tab:observations} using \emph{XMM} Science Analysis System
(\textsc{sas}) version 7.1.0, with the \textsc{rgsproc} pipeline. We
processed the 1st and 2nd order spectra, including 99 per cent of the
point spread function (PSF) and 97 per cent of the pulse-height
distribution, in order to get as many source photons as possible
without increasing background too much. We examined the observation
light-curves for CCD number 9 at absolute values of the cross
dispersion greater than $1.5\times10^{-4}$ (using values of FLAG of 8
and 16). The lightcurves were relatively consistent at values of
around $0.1 \ps$, except for some short flaring in the longer
observation. We filtered this observation, excluding time periods with
count rates greater than $0.2 \ps$.

Centaurus fills the entire field of view of the RGS instruments. We
therefore required blank-sky background spectra to subtract
instrumental and external backgrounds. As we used a 97~per~cent
pulse-height distribution cut we could not use that standard
\textsc{rgsbkgmodel} tool. Instead we selected five deep RGS
observations of point-like sources from relatively low Galactic
latitude to generate background spectra (Table
\ref{tab:observations}).  We processed these observations with the
same parameters as Centaurus, cleaning flares in the same way, and
excluding the inner 90 per~cent PSF (where the sources lie) to
generate the background spectra.

We combined with \textsc{rgscombine} the separate background
observations to make RGS 1 and RGS 2 spectra for the two spectral
orders. We used \textsc{rgscombine} to add the spectra and responses
from the two foreground Centaurus observations, after reprocessing
including the correct attitude values. The background spectra
contained some invalid spectral channels which did not correspond with
the foreground spectra, so we marked these channels as invalid in the
foreground spectra. We grouped the foreground spectra to have a
minimum of 25 counts per spectral bin.

We checked the background spectra were the same by applying them to
the foreground individually. The backgrounds were indistinguishable
in the 6.5 to 27~{\AA} range.
 
\begin{table*}
  \caption{\emph{XMM-Newton} RGS observations of the target
    (Centaurus) and the five background fields. RA and Dec are the
    boresight pointing coordinates, length specifies
    the observation length, and the exposure is the RGS1 exposure
    after cleaning for bad time periods (we total 158 and 157 ks
    exposure on the target in the RGS1 and 2, respectively).
    $N_\mathrm{H}$ shows the weighted Galactic Hydrogen column density
    towards the object \citep{Kalberla05}.}

  \begin{tabular}{llllllll}
    \emph{XMM} observation ID & Target & RA (J2000) & Dec (J2000) & Date & Length (ks) &
    Exposure (ks) & $N_\mathrm{H}$ ($10^{20} \psqcm$) \\ \hline
    0046340101 & Centaurus cluster & $12^\mathrm{h}48^\mathrm{m}47.93^\mathrm{s}$ & $-41^\circ18'43.5''$ & 2002-01-03 & 47.8 & 46.5 & 8.56 \\
    0406200101 & Centaurus cluster & $12^\mathrm{h}48^\mathrm{m}43.05^\mathrm{s}$ & $-41^\circ18'42.5''$ & 2006-07-25 & 124.3 & 111.7 & 8.56 \\
    0112210201 & NGC 3783 & $11^\mathrm{h}39^\mathrm{m}01.70^\mathrm{s}$ & $-37^\circ44'08.8''$ & 2001-12-17 & 137.8 & 72.5 & 9.91 \\
    0112210501 & NGC 3783 & $11^\mathrm{h}39^\mathrm{m}01.73^\mathrm{s}$ & $-37^\circ44'08.7''$ & 2001-12-19 & 137.8 & 124.2 & 9.91 \\
    0300430101 & NGC 3256 & $10^\mathrm{h}27^\mathrm{m}51.45^\mathrm{s}$ & $-43^\circ54'14.0''$ & 2005-12-06 & 134.0 & 116.6 & 9.14 \\
    0302850201 & MCG-5-23-16 & $09^\mathrm{h}47^\mathrm{m}45.43^\mathrm{s}$ & $-30^\circ56'57.6''$ & 2005-12-10 & 131.2 & 111.7 & 8.69 \\
    0305310101 & EC 13471-1258 & $13^\mathrm{h}49^\mathrm{m}52.17^\mathrm{s}$ & $-13^\circ13'36.0''$ & 2006-01-27 & 119.4 & 94.6 & 5.34 \\
  \end{tabular}
  \label{tab:observations}
\end{table*}

\begin{figure*}
  \includegraphics[width=0.8\textwidth]{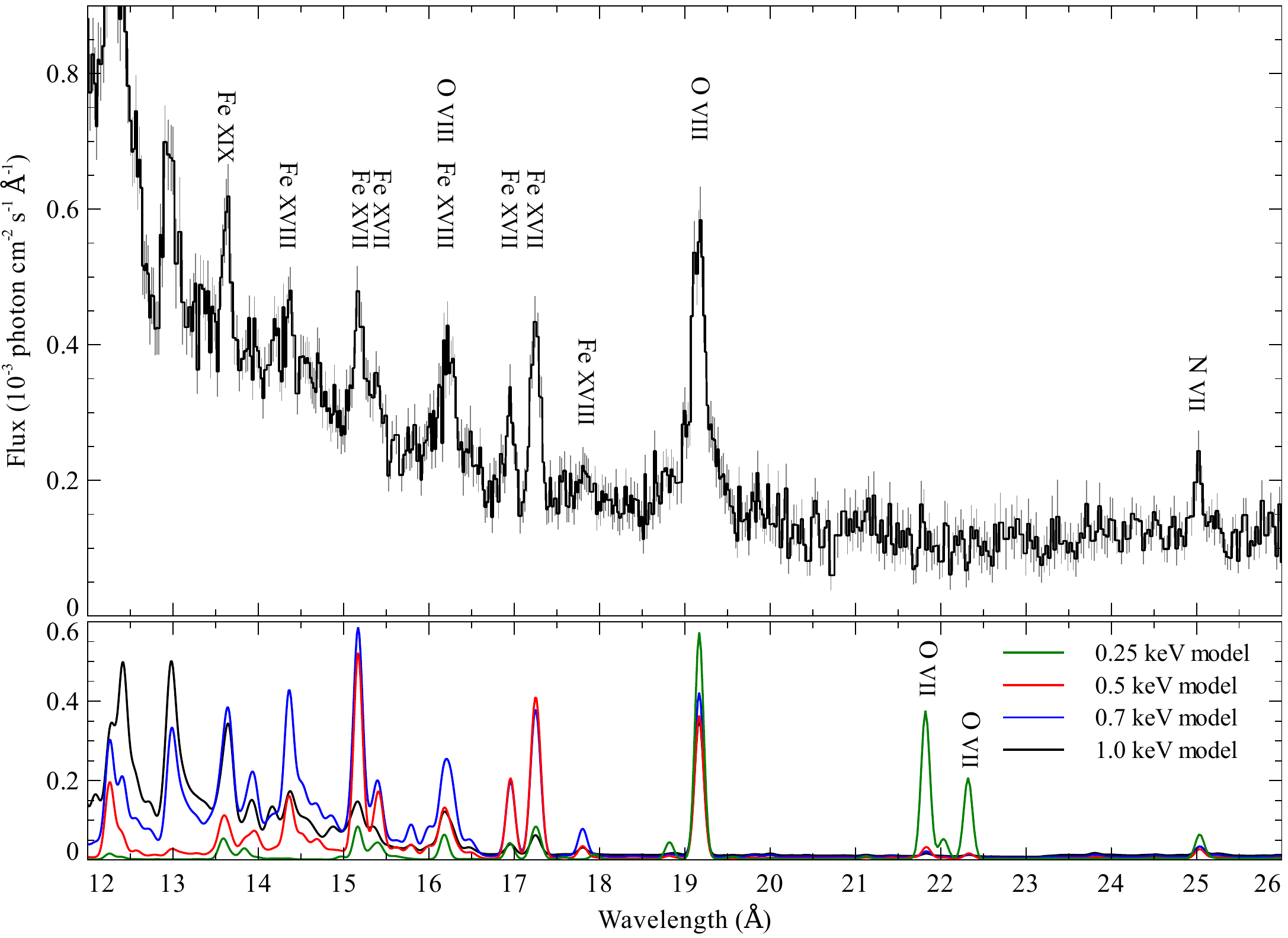}
  \caption{Top panel: fluxed and combined 1st and 2nd order RGS1 and
    RGS2 data, using 90 per~cent of the RGS PSF. The spectrum has been
    binned by a factor of 2 below 20{\AA} (giving 0.02{\AA} bins) and
    3 above that. Bottom panel: smoothed solar-abundance model
    \textsc{apec} spectra at 0.25, 0.5, 0.7 and 1 keV with arbitrary
    normalisation.}
  \label{fig:spectrumzoom}
\end{figure*}

\section{Spectrum}
We show in Fig.~\ref{fig:spectrum} the full fluxed spectrum using the
99~per~cent PSF with 97 per cent of the pulse-height distribution.  A
99 per~cent PSF corresponds to roughly the inner 160~arcsec width in
the cross-dispersion direction. This was created by combining the
first and second order spectra for both observations together with the
task \textsc{rgsfluxer}, subtracting the combined background
spectra. We do not show the spectrum at wavelengths longer than
26{\AA} as the background becomes increasingly important. We note that
the output from \textsc{rgsfluxer} is for display purposes only. We do
not use it to obtain quantitative information.

In the plot we also show the spectrum extracted from the inner 90
per~cent of the PSF (which corresponds to around 60~arcsec width in
the cross-dispersion direction). In addition we plot the difference
between the two spectra, which is equivalent to the spectrum extracted
between 60 and 160 arcsec in the cross-dispersion direction. In the
lower panel we show smoothed model \textsc{apec} \citep{SmithApec01}
spectra for plasmas with Solar metallicity at temperatures of 0.5, 0.7
and 1~keV for comparison. These models are plotted with arbitrary
normalisation to have roughly the same range in values. Note that the
spectrum also contains lines from gas hotter than 1~keV.

The spectrum shows a variety of emission lines, from N, O, Ne, Mg, Si
and Fe in several ionisation states. Most interesting are those lines
indicating cool X-ray emitting gas, particularly strong in the 90
per~cent PSF spectrum. Fe~L lines from Fe~\textsc{xxiv} down to
Fe~\textsc{xvii} are observed. The strong N~\textsc{vii} line, in
particular, is interesting.

We show a zoom-up of the spectrum between 12 to 26{\AA} from the
90~per~cent PSF in Fig.~\ref{fig:spectrumzoom}. The plot shows we
clearly observe several distinct Fe~\textsc{xvii} lines. These lines
are relatively narrow, indicating that they come from a relatively
small region (See Section \ref{sect:xdisp}). Absent from the spectra
is any evidence for O~\textsc{vii} emission (which would appear at
21.8 and 22.2{\AA} at this redshift), which is a strong indicator of
gas less than 0.2~keV.

\begin{figure*}
  \includegraphics[width=0.85\textwidth]{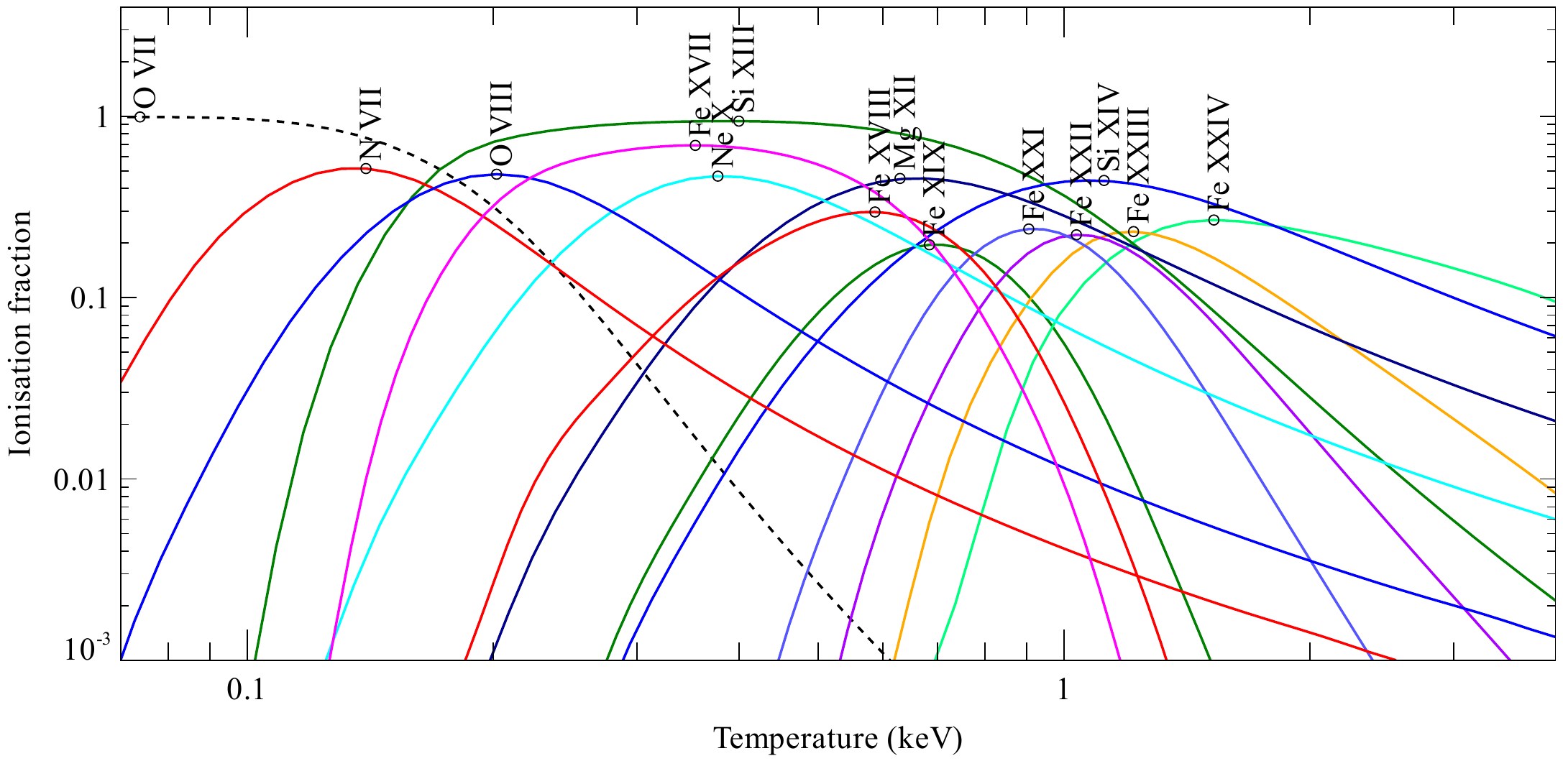}
  \caption{Ionisation equilibrium for lines from observed (solid) and
    unobserved (dotted) ions as function of temperature in keV
    {\citep{Mazzotta98}}. The data are shown here smoothed with cubic
    splines.}
  \label{fig:ioneq}
\end{figure*}

Different lines are sensitive to gas at different temperatures. The
strength of lines from a particular ion are governed by how much of an
element is in the form of that ion, the abundance of the element and
the density of the gas. We plot in Fig.~\ref{fig:ioneq} the fraction
of an element which is in the form of a particular ion as a function
of temperature, for those ions from which we observe lines (plus
O~\textsc{vii} which we do not observe). These results are from the
calculations of ionisation equilibrium of \cite{Mazzotta98} (which are
used by the \textsc{apec} model).

\begin{figure*}
  \includegraphics[width=0.99\textwidth]{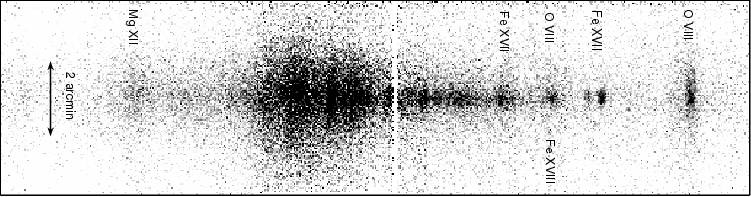}
  \caption{Image of the first order exposure-corrected dispersed
    spectrum, combining both RGS instruments and observations. The
    dispersion direction lies along the horizontal axis, with
    increasing wavelength rightwards. The vertical cross dispersion
    direction shows a one-dimensional image of the central $\sim
    5$~arcmin of the cluster.}
  \label{fig:xdispimage}
\end{figure*}

We show in Fig.~\ref{fig:xdispimage} the image formed by plotting the
dispersion angle of detected photons against the cross-dispersion
angle. Photons outside of the first order spectrum were removed by
making an energy-cut using the pulse-invariant CCD energy values. It
shows the combined RGS1 and RGS2 cross-dispersion image for the two
observations, exposure-corrected for bad pixels. The horizontal axis
shows increasing wavelength, while the off-axis angle in the
cross-dispersion direction is shown vertically. It can be seen that
some lines, e.g. Fe~\textsc{xvii}, are much more centrally concentrated
than others, e.g. O~\textsc{viii}.

\subsection{Line cross-dispersion profiles}
\label{sect:xdisp}
\begin{figure}
  \includegraphics[width=\columnwidth]{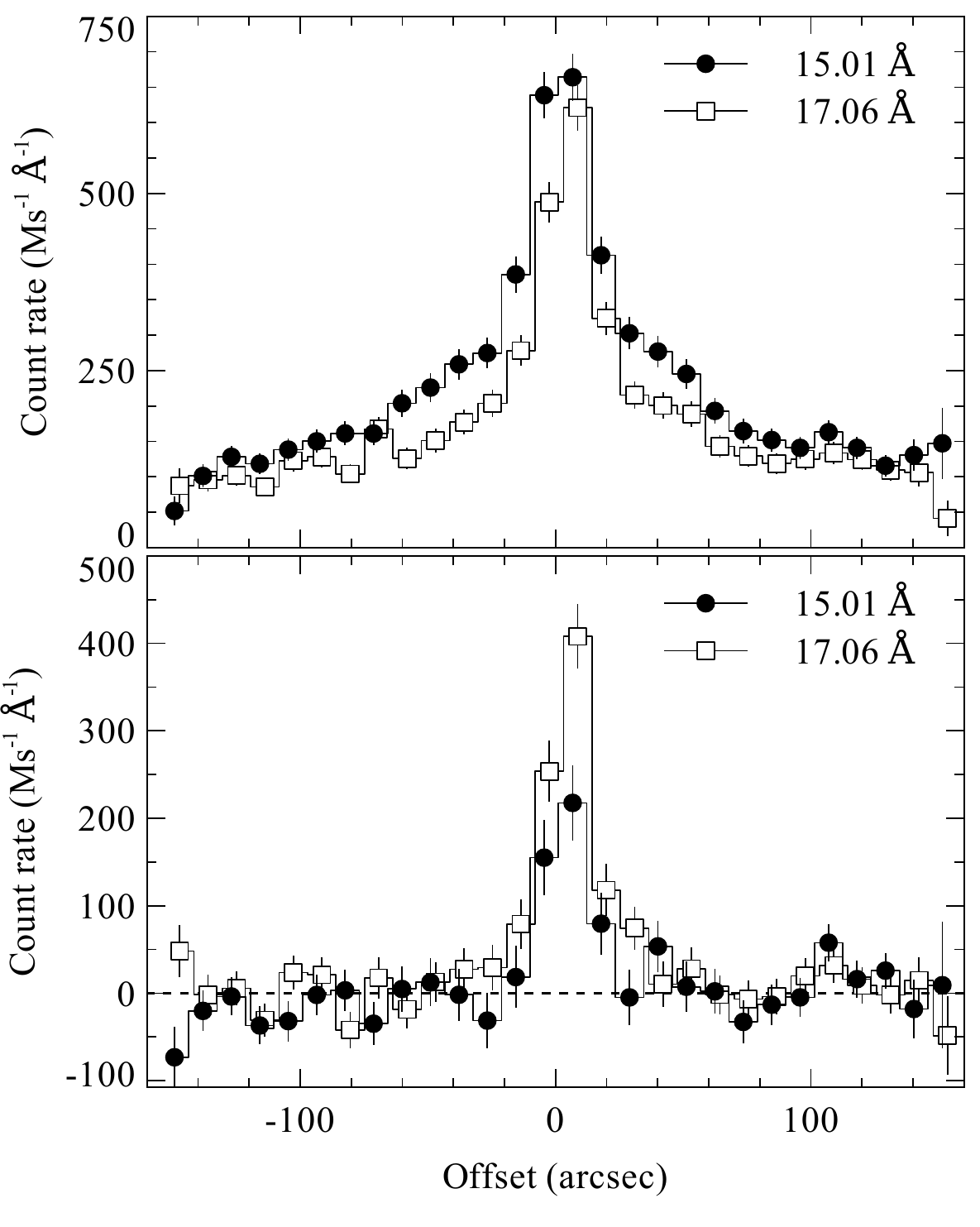}
  \caption{Line profiles as a function of cross-dispersion angle for
    the two strongest Fe~\textsc{xvii} lines. The top and bottom panel
    shows before and after subtracting neighbouring continuum,
    respectively.}
  \label{fig:lineprof_fe}
\end{figure}

We can examine this more quantitatively by plotting the profiles of
individual lines as a function of cross-dispersion angle. As one of
the most interesting cases, we show in Fig.~\ref{fig:lineprof_fe} the
profiles of the two strongest Fe~\textsc{xvii} lines.  The 15.01{\AA}
line can be strongly resonantly scattered as its oscillator strength
is 2.7.  Looking at the top panel of Fig.~\ref{fig:lineprof_fe}, which
shows the raw profile in a $\pm 0.1${\AA} strip either side of the
line centre, it appears there is evidence for scattering (a broader
15{\AA} profile). However, subtracting nearby continuum emission
creates much more sharply peaked distributions of emission (shown in
the bottom panel of Fig.~\ref{fig:lineprof_fe}), consistent between
the two lines. This is much more in line with the predictions from the
\emph{Chandra} Centaurus data of a resonance scattering model
\citep{SandersReson06}, which predicts at most a few per~cent effect
(resonant scattering is strong in the very central, cool, dense
regions, but this cannot scatter radiation to appear to come from a
larger radius).

Fitting a Gaussian to the continuum-subtracted profiles gives half
energy width (HEW) values of 14.2 and 12.3~arcsec for the 17.06{\AA}
and 15.01{\AA} lines, respectively.  These values are consistent with
the approximate HEW of the RGS mirrors (13.8 and 13.0 for the RGS1 and
RGS2, respectively).  The 17.06{\AA} blended line strength appears
substantially stronger than the 15.01{\AA} signal. We will reexamine
this in Section~\ref{sect:directline}.

\begin{figure}
  \includegraphics[width=\columnwidth]{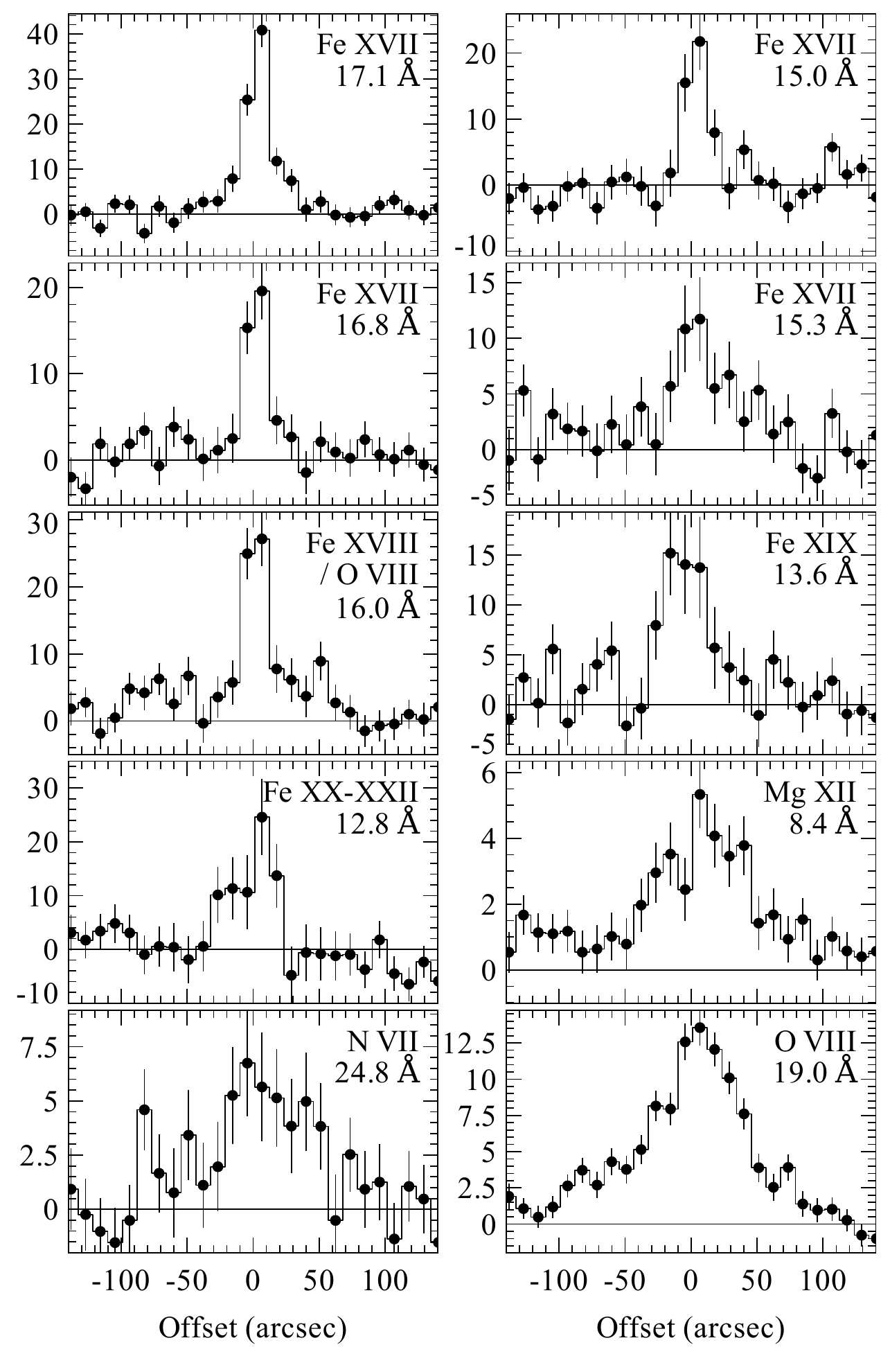}
  \caption{Profiles of several prominent lines in the cross-dispersion
    direction. The units are in $10^{-5} \: \mathrm{counts} \ps
    $\AA$^{-1}$. Continuum has been subtracted from a neighbouring
    part of the spectrum.}
  \label{fig:lineprof_others}
\end{figure}

We show in Fig.~\ref{fig:lineprof_others} cross-dispersion profiles of
several other strong lines. In each of these cases we plot the profile
0.1{\AA} either side of the centre (except for the broad
O~\textsc{viii} and Mg~\textsc{xii} lines, where we use 0.2{\AA}). The
general picture is that lines that are emitted by cooler gas
(e.g. Fe~\textsc{xvii}, \textsc{xviii}, \textsc{xix}) come from more
compact regions. One line which differs from this picture is the
N~\textsc{vii} line, which is fairly broad in the cross-dispersion
profile. Fig.~\ref{fig:spectrum} however shows that most of the
contribution to this line comes from the inner 90 per~cent of the PSF.

\section{Spectral fitting}
\begin{table*}
  \caption{Spectral fit results to the data extracted from the
    99~per~cent PSF aperture. If two or more results are given,
    this means the parameter was allowed to vary independently in each
    component, otherwise they were tied together. Upper limits are $2
    \sigma$. For the multi-temperature models the number in brackets
    signify the temperature component number the best fitting
    parameters apply to. The values of the source sizes are HEW,
    exclusive of the 13.2 arcsec component from the mirrors.}

  \begin{tabular}{l|llllll}
    Model &
    \textsc{vapec+vapec} &
    \parbox[t]{0.8in}{\textsc{vapec+vmcflow}\\k$T_\mathrm{min} = 0.08 \keV$} &
    \parbox[t]{0.8in}{\textsc{vapec+vmcflow}\\k$T_\mathrm{min}$ free} &
    \textsc{$5\times$vapec} &
    \textsc{$5\times$vmcflow} &
    %\textsc{$8\times$vapec} &
    \\ \hline
    k$T$ (keV) &
    \parbox[t]{0.8in}{$1.85 \pm 0.02$\\$0.766 \pm 0.008$} & % vapec+vapec 
    $1.86 \pm 0.02$ & % vapec+vmcflow
    \parbox[t]{0.8in}{$2.04 \pm 0.02$\\$0.54 \pm 0.01$} & % vapec+vmcflowmin
    \parbox[t]{0.8in}{3.2, 2.4, 1.6, 0.8, 0.4} & % 5vapec
    \parbox[t]{0.8in}{3.2-2.4, 2.4-1.6, 1.6-0.8, 0.8-0.4, 0.4-0.0808} & % 5vmcflow
    %\parbox[t]{0.8in}{4.32, 2.72, 1.72, 1.37, 1.08, 0.86, 0.68 and 0.34} & % 8vapec
    \\
    $N_\mathrm{H}$ ($10^{20} \psqcm$) &
    $10.1 \pm 0.2$ & % vapec+vapec
    $10.1 \pm 0.2$ & % vapec+vmcflow
    $9.7 \pm 0.2$ & % vapec+vmcflowmin
    $9.0 \pm 0.2$ & % 5vapec
    $8.8 \pm 0.2$ & % 5mkcflow
    %$8.6 \pm 0.2$ & % 8vapec
    \\
    sizes (arcmin) &
    \parbox[t]{0.8in}{$1.12 \pm 0.03$\\$0.32 \pm 0.03$} & % vapec+vapec
    \parbox[t]{0.8in}{$1.19 \pm 0.03$\\$0.37 \pm 0.03$} & % vapec+vmcflow
    \parbox[t]{0.8in}{$1.42 \pm 0.04$\\$0.35 \pm 0.02$} & % vapec+vmcflowmin
    \parbox[t]{0.8in}{$1.96 \pm 0.09$ (1-2), $0.56 \pm 0.03$
      (3), $0.39 \pm 0.04$ (4-5)} & % 5vapec
    \parbox[t]{0.8in}{$\sim 3$ (1), $0.99 \pm 0.04$ (2), $0.32 \pm
      0.06$ (3), $0.49 \pm 0.09$ (4-5)} & % 5vmcflow
    %\parbox[t]{0.8in}{$\sim 3$ (1-2), $0.80 \pm 0.3$ (3-5), $0.38
    %  \pm 0.03$ (6-8)} & % 8vapec
    \\
    N &
    \parbox[t]{0.8in}{$<0.18$\\$3.8^{+0.5}_{-0.8}$} & % vapec+vapec
    \parbox[t]{0.8in}{$<0.18$\\$0.86 \pm 0.12$} & % vapec+vmcflow
    \parbox[t]{0.8in}{$<0.21$\\$2.0 \pm 0.3$} & % vapec+vmcflowmin
    \parbox[t]{0.8in}{$<0.15$ (1-2)\\$1.6 \pm 0.2$ (3-5)} & % 5vapec
    \parbox[t]{0.8in}{$<0.18$ (1-2)\\$3.5 \pm 0.6$ (3-5)} & % 5vmkcflow
    %\parbox[t]{0.8in}{$<0.17$ (1-5)\\$5.2 \pm 0.9$ (6-8)} & % 8vapec
    \\
    O &
    \parbox[t]{0.7in}{$0.32 \pm 0.02$\\$0.41 \pm 0.09$} & % vapec+vapec
    \parbox[t]{0.7in}{$0.34 \pm 0.02$\\$0.15 \pm 0.03$} & % vapec+vmcflow
    \parbox[t]{0.7in}{$0.40 \pm 0.03$\\$0.34 \pm 0.05$} & % vapec+vmcflowmin
    $0.47 \pm 0.02$ & % 5vapec
    $0.52 \pm 0.02$ & % 5vmkcflow
    %$0.51 \pm 0.02$ & % 8vapec
    \\
    Ne &
    $0.45 \pm 0.03$ & % vapec+vapec
    $0.37 \pm 0.02$ & % vapec+vmcflow
    $0.56 \pm 0.04$ & % vapec+vmcflowmin
    $0.74 \pm 0.04$ & % 5vapec
    $0.80 \pm 0.05$ & % 5vmkcflow
    %$0.80 \pm 0.04$ & % 8vapec
    \\
    Mg &
    $0.60 \pm 0.03$ & % vapec+vapec
    $0.49^{+0.03}_{-0.02}$ & % vapec+vmcflow
    $0.69 \pm 0.05$ & % vapec+vmcflow2
    $0.91 \pm 0.05$ & % 5vapec
    $1.02 \pm 0.07$ & % 5mkcflow
    %$1.03 \pm 0.05$ & % 8vapec
    \\
    Si &
    $1.22 \pm 0.06$ & % vapec+vapec
    $1.02 \pm 0.06$ & % vapec+vmcflow
    $1.34 \pm 0.05$ & % vapec+vmcflowmin
    $1.60 \pm 0.09$ & % 5vapec
    $1.76 \pm 0.12$ & % 5mkcflow
    %$1.70 \pm 0.10$ & % 8vapec
    \\
    Ca &
    $2.0 \pm 0.3$ & % vapec+vapec
    $2.2 \pm 0.3$ & % vapec+vmcflow
    $2.1 \pm 0.2$ & % vapec+vmcflowmin
    $2.6 \pm 0.4$ & % 5vapec
    $2.8 \pm 0.5$ & % 5mkcflow
    %$2.5 \pm 0.5$ & % 8vapec
    \\
    Fe &
    \parbox[t]{0.7in}{$0.66 \pm 0.02$\\$0.53 \pm 0.04$} & % vapec+vapec
    \parbox[t]{0.7in}{$0.67 \pm 0.02$\\$0.29 \pm 0.03$} & % vapec+vmcflow
    \parbox[t]{0.7in}{$0.82 \pm 0.03$\\$0.59 \pm 0.07$} & % vapec+vmcflowmin
    $0.97 \pm 0.02$ & % 5vapec
    $1.08 \pm 0.02$ & % 5mkcflow
    %$1.04 \pm 0.05$ & % 8vapec
    \\
    Ni &
    $1.81 \pm 0.08$ & % vapec+vapec
    $1.67 \pm 0.07$ & % vapec+vmcflow
    $1.95 \pm 0.09$ & % vapec+vmcflowmin
    $2.34 \pm 0.11$ & % 5vapec
    $2.54 \pm 0.15$ & % 5mkcflow
    %$2.52 \pm 0.12$ & % 8vapec
    \\
    norm &
    \parbox[t]{1in}{$(4.14 \pm 0.04) \times 10^{-2}$\\
      $(2.0 \pm 0.1) \times 10^{-3}$} & % vapec+vapec
    $(3.89 \pm 0.07) \times 10^{-2}$ & % vapec+vmcflow
    $(3.46 \pm 0.06) \times 10^{-2}$ & % vapec+vmcflowmin
    \parbox[t]{1in}{See Fig.~\ref{fig:norms}} & % 5vapec
    -- & % 5mkcflow
    %\parbox[t]{1in}{See Fig.~\ref{fig:norms}} & % 8vapec
    \\
    $\dot{M}$ ($\Msunpyr$) &
    -- & % vapec+vapec
    $7.4 \pm 0.5$ & % vapec+vmcflow
    $8.7 \pm 0.5$ & % vapec+vmcflowmin
    -- & % 5vapec
    \parbox[t]{1in}{See Fig.~\ref{fig:mdott}} % 5mkcflow
    %-- & % 8vapec
    \\
    $\chi_\nu^2$ &
    $6435/5068 = 1.27$ & % vapec+vapec
    $6494/5069 = 1.28$ & % vapec+vmcflow
    $6381/5068 = 1.26$ & % vapec+vmcflowmin
    $6362/5069 = 1.26$ & % 5vapec
    $6397/5069 = 1.26$ & % 5vmcflow
    %$6361/5067 = 1.26$ & % 8vapec
    \\
  \end{tabular}
  \label{tab:fitresults}
\end{table*}

The observed spectrum is affected by several factors, listed below.
\begin{enumerate}
\item The temperature of the observed region affects the ratio of the
  strengths of the emission lines and the continuum. Emission lines
  are increasingly important below $\sim 1$~keV. Some lines,
  e.g. Fe~\textsc{xvii}, are only emitted over certain restricted
  temperature ranges (see Fig.~\ref{fig:ioneq}). The spectrum will be
  the sum of several temperature components which means care must be
  taken in using line ratios as temperature diagnostics.

\item The metallicity of a temperature component affects the strength
  of its emission lines. In general the metallicity of each element in
  the plasma can vary as a function of temperature and position.

\item The spatial size of the emitting region causes spectral
  broadening. If a temperature component comes from a smaller spatial
  region, its lines will be broadened less than a component from a
  large region. Lines are broadened by \citep{Brinkman98}
  \begin{equation}
    \Delta \lambda = \frac{0.124}{m} \Delta \theta \: \textrm{\AA},
    \label{eqn:smoothing}
  \end{equation}
  where $\Delta \theta$ is the source extent (HEW; half energy width)
  in arcmin, and $m$ is the spectral order. The \emph{XMM} RGS
  response already takes account of the broadening for point sources,
  so this width is on top of the $\sim 13.2$~arcsec HEW PSF of the
  mirror modules of the RGS.

\item The sensitivity of the RGS is a function of the position on the
  sky. It is less sensitive to components observed from off axis.
\end{enumerate}
Therefore a complex source like the core of the Centaurus cluster
would require a significant amount of modelling to account for each of
these effects. There are also degeneracies when fitting the
spectra. For example, the emission measure of low temperature gas is
inversely proportional to its metallicity as the continuum is
difficult to measure.

Our general technique in this paper is to fit multiple thermal models
to the spectrum extracted from the RGS. Each thermal model is smoothed
by a variable-width Gaussian (of width in wavelength given in equation
\ref{eqn:smoothing}) to take account of the spatial distribution
(which assumes the emission region is Gaussian in shape, centred on
the cluster centre). We tie the abundances of each of the thermal
components together, unless the data are good enough to constrain them
in individual components. The thermal components are absorbed by
Galactic absorption before smoothing with the Gaussian model.

Our modelling assumes that each component has a fixed temperature,
metallicity and spatial scale. In most of the spectral fits we make
assumptions in our choice of which parameters to vary. For example, we
do not allow many of the metallicities to vary with temperature.

We fitted the first-order spectra between 6.5 and 27\AA, where the
background is low relative to the foreground. The second-order spectra
were fitted between 7 and 15.5\AA. \textsc{xspec} version 12.3.1
\citep{ArnaudXspec} was used to do the spectral fitting. We use the
spectra extracted from the 99~per~cent PSF to get as complete a
picture as possible.

\subsection{Single temperature model}
\begin{figure*}
  \includegraphics[width=0.9\textwidth]{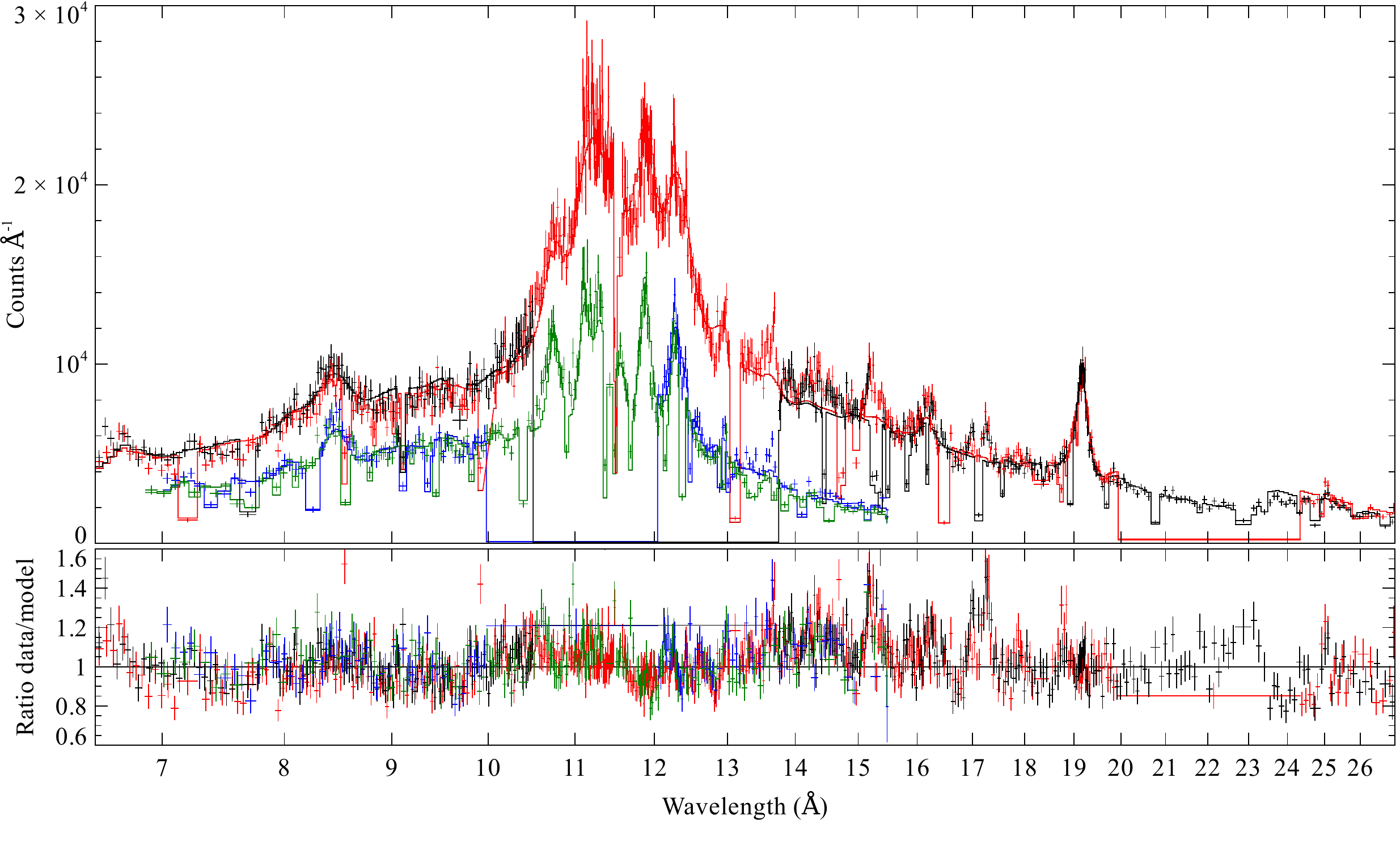}
  \caption{Best fitting single-temperature model and ratio of data to
    model for the 99~per~cent PSF spectra. The points are the first
    and second order data from the two RGS detectors and the lines are
    the best fitting model after folding through the response. The
    data have been rebinned for display in \textsc{xspec} to have a
    signal to noise ratio of 13. The model fails to account for the
    Fe~\textsc{xvii} and N~\textsc{vii} lines.}
  \label{fig:singletemp}
\end{figure*}

To demonstrate that an isothermal model is inadequate, we show in
Fig.~\ref{fig:singletemp} the best fitting single temperature
\textsc{vapec} \citep{SmithApec01} model. In this model the
temperature, Galactic column density and normalisation are free. The
relative normalisations of each RGS dataset are allowed to vary
relative to the RGS1 first order spectrum. The N, O, Ne, Mg, Si, Ca,
Fe and Ni metallicities are free to vary. The spectrum is smoothed by
a variable-width Gaussian to account for spatial broadening of the
spectral lines (Equation \ref{eqn:smoothing}).  The 1.6~keV model
fails to account for the Fe~\textsc{xvii} and N~\textsc{vii} lines
($\chi^2_\nu = 7474/5075 = 1.47$).

\subsection{Simple two-temperature model}
The most simple extension of a single thermal component model is one
with two temperature components. This model consists of two
\textsc{vapec} components, each with a variable temperature,
normalisation, and N, O and Fe metallicities. They are absorbed by the
same Galactic column density, but are allowed to vary independently in
spatial size (smoothing according to Equation
\ref{eqn:smoothing}). The Ne, Mg, Si, Ca and Ni metallicites are tied
between the two components and allowed to vary independently. We allow
the N metallicity to vary between the two components because the
N~\textsc{vii} line is narrow. If the hot component is forced to have
the same N metallicity as the cool component, then the model cannot
fit the narrow line because too much emission comes from larger
spatial regions. Table~\ref{tab:fitresults} shows the best fitting
values of each of the parameters and their uncertainties. The
improvement in $\chi^2$ is around 1000 over the single temperature
model. The best fitting temperatures, 0.77 and 1.85~keV, are very
close to the range of values found in the inner arcmin by
\emph{Chandra} \citep{Fabian05}.

The best fitting iron metallicities, however, are substantially lower
than the $1.5-2 \Zsun$ peak values from \emph{Chandra} and \emph{XMM}
CCD spectra \citep{SandersEnrich06}. In fact all of the metallicities
(except Ca) of each of the elements we measure from the RGS spectra
are significantly lower than the CCD results. CCD measurements
indicate that the metallicities drops dramatically in the innermost
region \citep{SandersCent02}, which may correspond with this low
metallicity. We will discuss this further in
Section~\ref{sect:metals}.

The lines from cooler temperature gas are narrower than the lines from
hotter gas, leading to a larger spatial scale for the hot component
(1.1 vs 0.3~arcmin). The emission measure of the cooler component is
only around 5~per~cent of the hot component.

\subsection{Multi-temperature model}
\begin{figure}
  \includegraphics[width=\columnwidth]{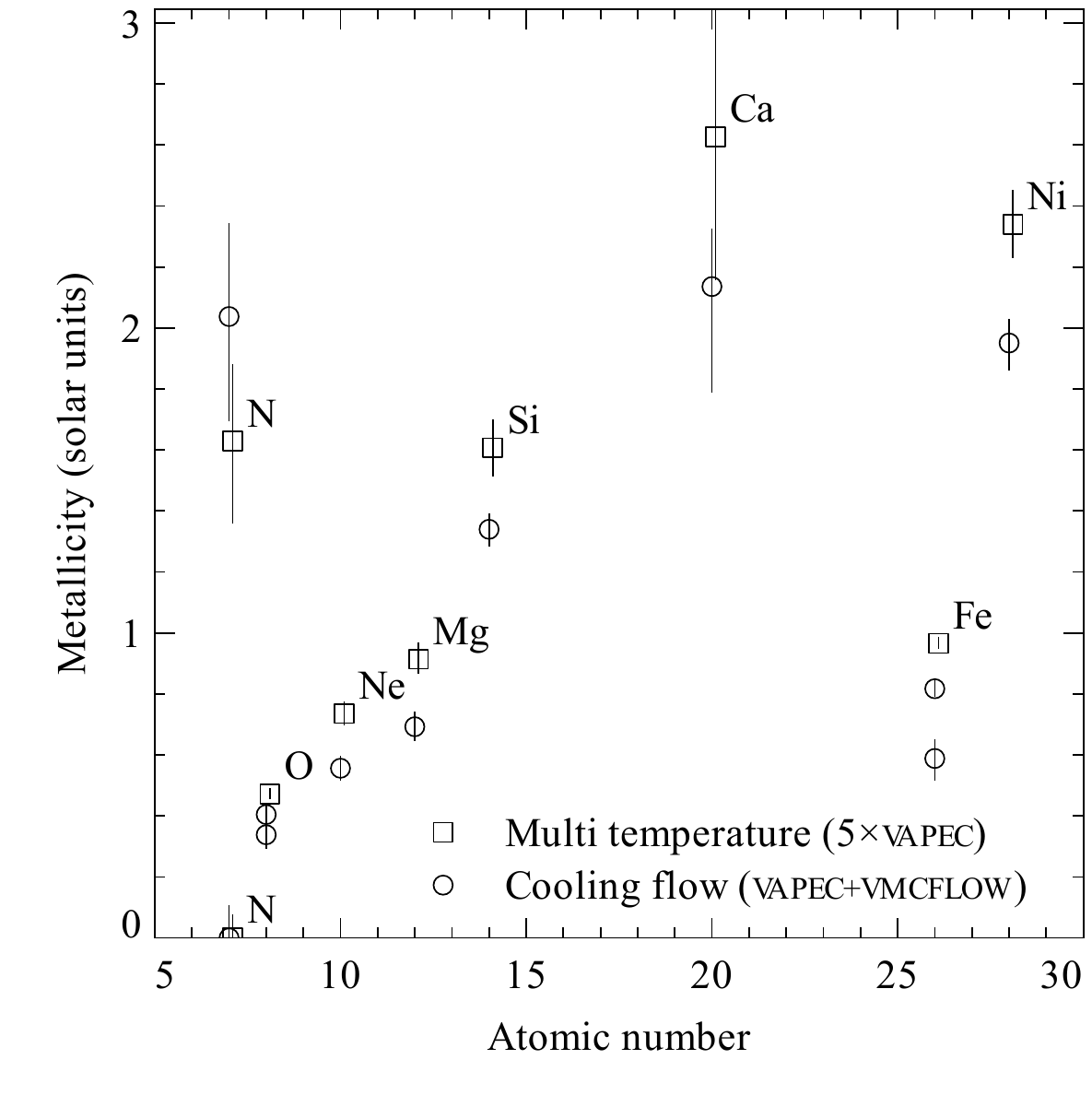}
  \caption{Best fitting metallicities for the spectral fits from
    multi-temperature ($5\times$\textsc{vapec}) and cooling flow model
    fits (\textsc{vapec}+\textsc{vmcflow} with free minimum
    temperature) to the data extracted from the 99~per~cent PSF. Note
    that N, O and Fe are allowed to vary between the two components in
    the \textsc{vapec}+\textsc{vmcflow} model.}
  \label{fig:metals}
\end{figure}

\begin{figure}
  \includegraphics[width=\columnwidth]{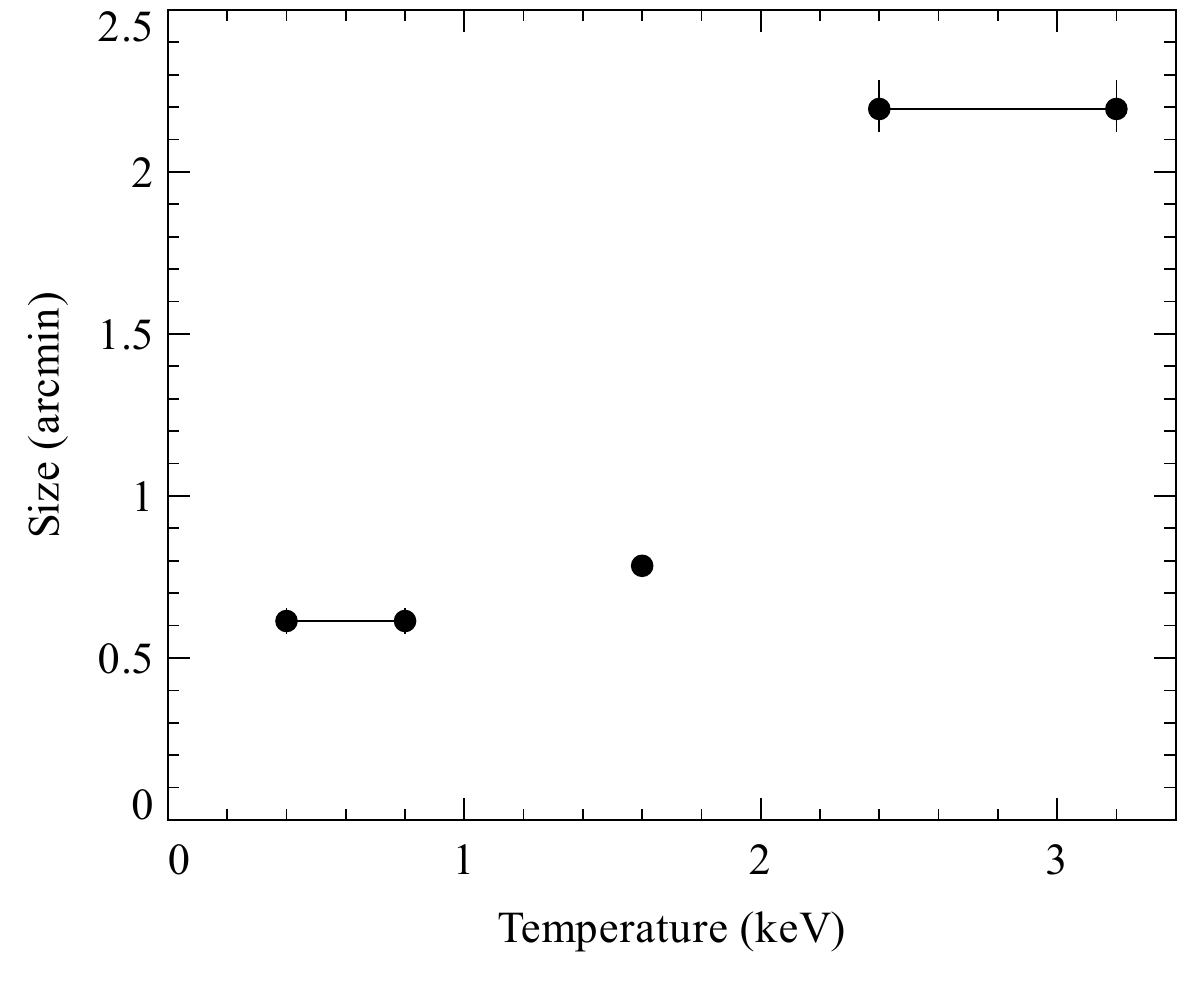}
  \caption{Best fitting spatial size of spectral fits using the line
    widths for the $5\times$\textsc{vapec} model (adding the PSF
    contribution). The lines connect temperature components which have
    their sizes tied together.}
  \label{fig:sizesT}
\end{figure}

\begin{figure}
  \includegraphics[width=\columnwidth]{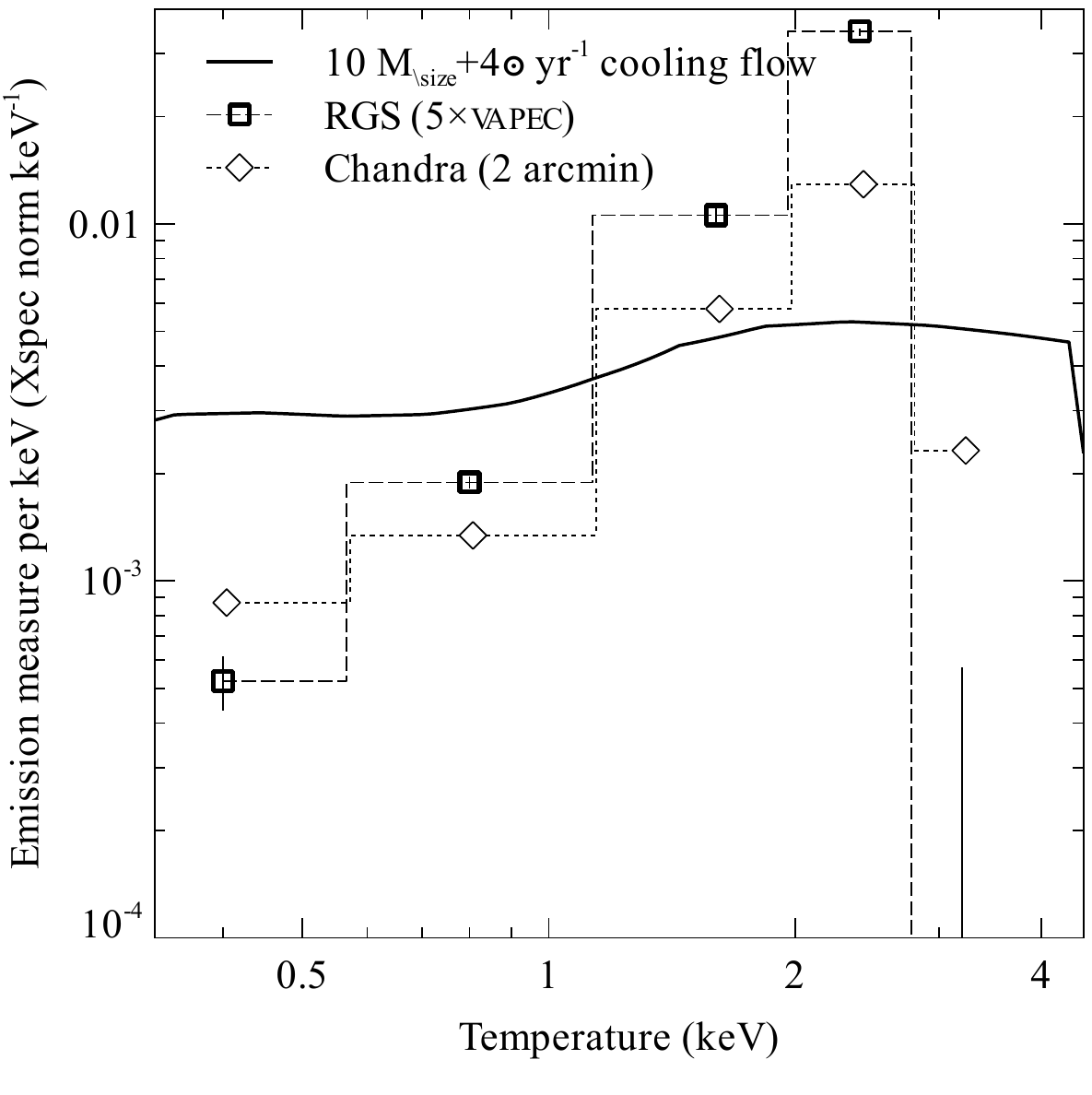}
  \caption{Comparison of the RGS with \emph{Chandra} emission measures
    per unit temperature as a function of temperature. The
    \emph{Chandra} results were extracted from a 2~arcmin radius
    around the core of the cluster. The solid line shows the expected
    distribution of emission measure for a $10 \Msunpyr$ isobaric
    cooling flow at Solar metallicity, cooling from a temperature of
    4.5~keV to 0.0808~keV using the \textsc{mkcflow} model.}
  \label{fig:norms}
\end{figure}

\begin{figure}
  \includegraphics[width=\columnwidth]{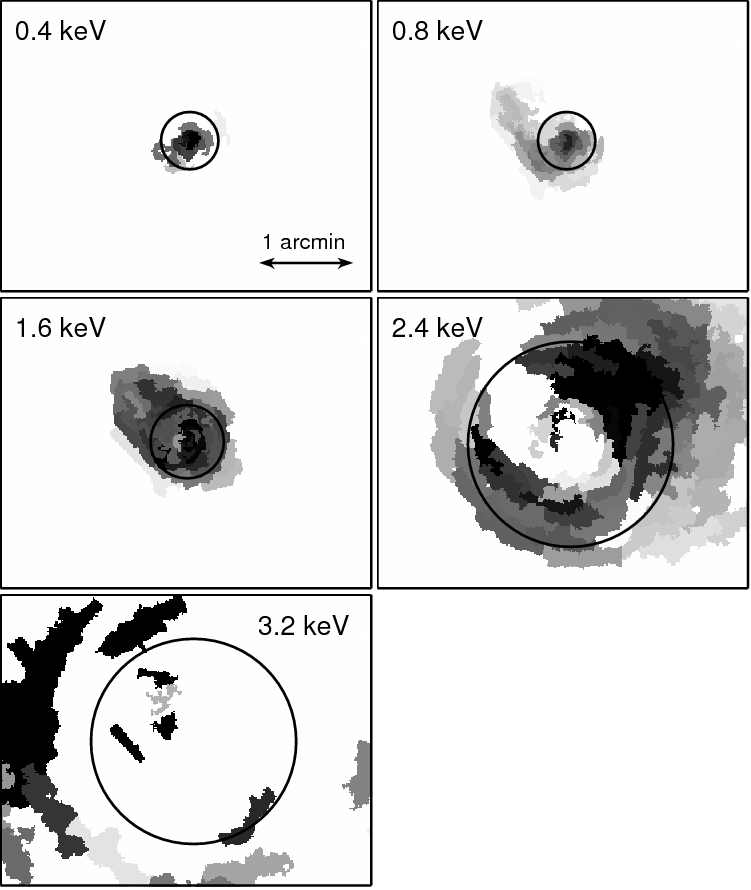}
  \caption{Maps of the \emph{Chandra} emission measure per unit area
    using multi temperature model. The circles show the best fitting
    HEW of the emission from the line widths in the spectral fits to
    the RGS data (the 0.4 and 0.8 keV components are forced to be
    equal, as are the 2.4 and 3.2 keV components), including the
    effect of the PSF. The RGS cross-dispersion direction for the
    longer observation lies along the cool plume.}
  \label{fig:sizemaps}
\end{figure}

We can extend the two-temperature model with more temperature
components. Multitemperature fits with several free temperatures
typically become unstable when spectral fitting. To avoid this we
constructed a model containing components with fixed temperatures,
allowing the emission measure of each component to vary.  We used a
multi-temperature model containing five components
($5\times$\textsc{vapec}). We used a range of temperatures in the
model to account for the ranges of temperatures in the cluster the
spectrum is sensitive to, with components at 0.4, 0.8, 1.6, 2.4 and
3.2 keV. We also tried an 8 component model, but the quality of the
fit was only slightly improved and it was difficult to constrain the
parameters.

Table~\ref{tab:fitresults} show the best fitting parameters for these
two models. To reduce the number of free parameters, we only used a
single set of metallicities for the temperature components (tieing O,
Ne, Mg, Si, Ca, Fe and Ni). We however split the components by
temperature to allow for two N values, otherwise there are obvious
residuals around the N~\textsc{vii} line. Rather than allow the
spatial scale of each of the components to vary, or fixing them to be
all the same, we apply three different spatial smoothing scales to
groups of the temperature components.

We plot the abundance of each variable relative to solar in
Fig.~\ref{fig:metals} for the $5\times$\textsc{vapec} model. In
Fig.~\ref{fig:sizesT} is shown the best fitting HEW sizes from the
line widths against the temperature of the component. We note that
velocity broadening of the lines could mimic spatial broadening,
however, velocities above $700\kmps$ are required to make a
substantial difference to these measurements.

The emission measure of each component per unit temperature (assuming
a temperature bin size which is half the logarithmic change to the
next adjacent temperature value) is plotted as a function of
temperature in Fig.~\ref{fig:norms}. In this plot we also show values
from a \emph{Chandra} spatially-resolved analysis of the inner
2~arcmin. These were created from spectra extracted from
contour-binned \citep{SandersBin06} regions containing around $\sim
10^4$ counts\footnote{These are the same regions and spectra used to
  create figures 6 and 7 in \cite{SandersEnrich06}, extracted from
  \emph{Chandra} observation IDs 504, 504, 4954, 4955 and 5310.}. The
spectra were fit with a multi-temperature $5 \times$\textsc{apec}
model with the same five fixed temperatures as the RGS fit, but
assuming Solar abundance ratios. The model assumes each temperature
component in individual regions has the same metallicity.

The RGS and \emph{Chandra} results are roughly similar, considering we
are comparing the \emph{Chandra} results in the inner 2~arcmin radius
with the larger field of view of the RGS instruments and the decline
of the RGS effective area off-axis. The one major discrepancy is the
apparent lack of any gas above 3~keV in the RGS results (see
Section~\ref{sect:metals} for a discussion). We also plot the HEW
size from the line widths on the \emph{Chandra} emission measure maps
on Fig.~\ref{fig:sizemaps}, showing they are comparable.

The solid line in Fig.~\ref{fig:norms} shows the expected distribution
of emission measure for a simple isobaric cooling flow, cooling at the
rate of $10 \Msunpyr$ without any heating. It can be seen that the
observed emission measure of gas decreases more steeply with
temperature than predicted by a simple cooling flow.

\subsection{Single cooling flow component}
We have investigated how well the observed spectrum can be modelled by
a simple cooling flow. Our first model was based on a \textsc{vapec}
component plus a \textsc{vmcflow} cooling flow component. We used the
new feature in \textsc{xspec} version 12 to base the cooling flow
model spectrum on a \textsc{apec} thermal model rather than a
\textsc{mekal} one. We note, however, that this form of the model is
not internally self consistent, as the quantity of gas at each
temperature is computed by assuming the luminosities of the
\textsc{mekal} model. We made our own consistent version of the model,
but this had no effect on the predicted spectrum, so we show results
from the \textsc{xspec} model here.

We tried two forms of the model: a full cooling flow where the lower
temperature of the cooling flow was constrained to be the minimum
possible (0.0808 keV) and the second reduced model where it was
allowed to be a free parameter. In this model we allowed the N, O and
Fe metallicites to vary between the \textsc{vapec} and
\textsc{vmcflow} components, but fixed the other metallicities to have
the same values in the two components.

The best fitting parameters for the two models are shown in
Table~\ref{tab:fitresults}. The full cooling flow model gives a mass
deposition rate of $7.4 \pm 0.5 \Msunpyr$, whereas the reduced model
obtained a rate of $8.7 \pm 0.5 \Msunpyr$ cooling to $0.54 \pm 0.01
\keV$. The metallicities of the cooling flow component were lower for
the model where the gas cools to the minimum value, presumably to
decrease the strength of the emission lines.

The reduced cooling flow model gives a substantially better quality of
fit to the spectrum than the full model ($\chi^2=6381$ versus 6494),
but is poorer than the five component multitemperature model
($\chi^2=6362$).

\subsection{Multiple cooling flow components}
\begin{figure}
  \includegraphics[width=\columnwidth]{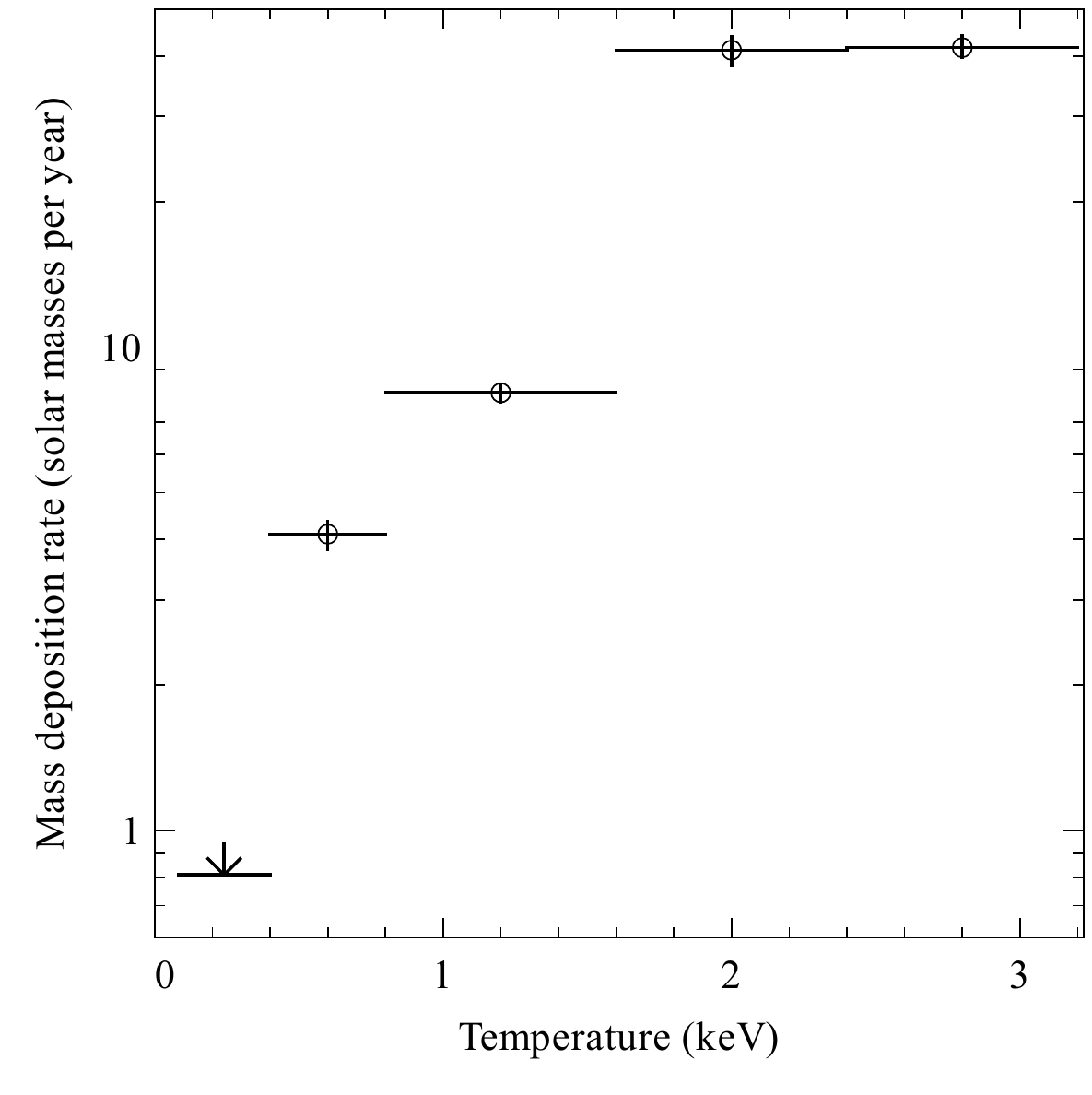}
  \caption{Mass deposition rates in the absence of heating using a
    cooling flow model made up of five different temperature ranges
    ($5\times$\textsc{vmcflow}).}
  \label{fig:mdott}
\end{figure}

A simple cooling flow is unlikely to be a good model to the complex
distribution of gas in the core of the Centaurus cluster, even in the
presence of cooling gas. We have therefore used a model where the
cooling flow is split into different temperature ranges and we allow
the mass deposition rate to vary in each range. We use fixed ranges in
temperature to obtain a stable fit.  The model examined here has steps
in temperature from 3.2 to 2.4, 2.4 to 1.6, 1.6 to 0.8, 0.8 to 0.4 and
0.4 to 0.0808~keV. The metallicities are assumed to be the same in
each component, except for the N metallicity in the three coolest
components, which is allowed to vary separately from the hotter
components.

We show in Fig.~\ref{fig:mdott} the mass deposition rate in the
absence of heating for each of the cooling flow components. We measure
mass deposition rates down to 0.4~keV and an upper limit of
$0.8\Msunpyr$ below that temperature. The results are consistent with
the multitemperature model results. There is systematically less gas
detected at lower temperature than expected from a simple cooling flow
(as in Fig.~\ref{fig:norms}).

\subsection{Direct spectral line measurements}
\label{sect:directline}
\begin{table*}
  \caption{Line luminosities for some of the more interesting lines.
    The listed wavelengths are rest
    wavelengths, averaging the position of lines in blends. The
    temperature range is the approximate range in temperature of gas
    the lines are emitted from (where the emissivity is within an
    order of magnitude of the peak). The widths are measurements of the Gaussian width,
    upper limits ($2 \sigma$) or
    fixed at 0.  The power and mass deposition rates are corrected for
    Galactic absorption. The mass deposition
    rates are calculated assuming a cooling flow cooling to 0.0808 from 4.5~keV
    temperature with Solar metallicities.}
  \begin{tabular}{llllll}
    Line & Wavelength (\AA)  & Temperature range (keV) & Width (\AA) & Power ($10^{39} \ergps$) & $\dot{M}$ (\Msunpyr) \\ \hline
    Fe \textsc{xvii} & 15.01 & $0.2 \rightarrow 0.9$ & $<0.07$       & $12.4 \pm 1.5$       & $1.6$ \\
    Fe \textsc{xvii} & 15.26 & $0.2 \rightarrow 0.9$ & $0$           & $3.6 \pm 1.3$        & $1.6$ \\
    Fe \textsc{xvii} & 16.78 & $0.2 \rightarrow 0.9$ & $0$           & $4.3 \pm 1.2$        & $1.7$ \\
    Fe \textsc{xvii} & 17.06 & $0.2 \rightarrow 0.9$ & $0.047 \pm 0.009$ & $19.3 \pm 2.0$   & $3.0$ \\
    N \textsc{vii}   & 24.78 & $ < 0.85 $            & $0.050 \pm 0.014$ & $11.5 \pm 1.6$   & $9.7$ \\
    O \textsc{vii}   & 21.60 & $ < 0.4  $            & $0$           & $<4.6$               & $<3.4$ \\
    O \textsc{vii}   & 22.10 & $ < 0.4  $            & $0$           & $<7.7$               & $<2.9$ \\
  \end{tabular}
  \label{tab:mdotlines}
\end{table*}

Instead of direct spectral fitting to the whole of the spectrum, the
strength of the emission lines can be used to gauge the amount of
cooling taking place through the temperature range they are sensitive
to. The amount of flux in a line can be compared to that expected from
a cooling flow model.  As there are several possible lines which can
be examined, this allows independent determinations of the mass
deposition rate, but does not use the full spectral information
available when spectral fitting.

\begin{figure}
  \includegraphics[width=\columnwidth]{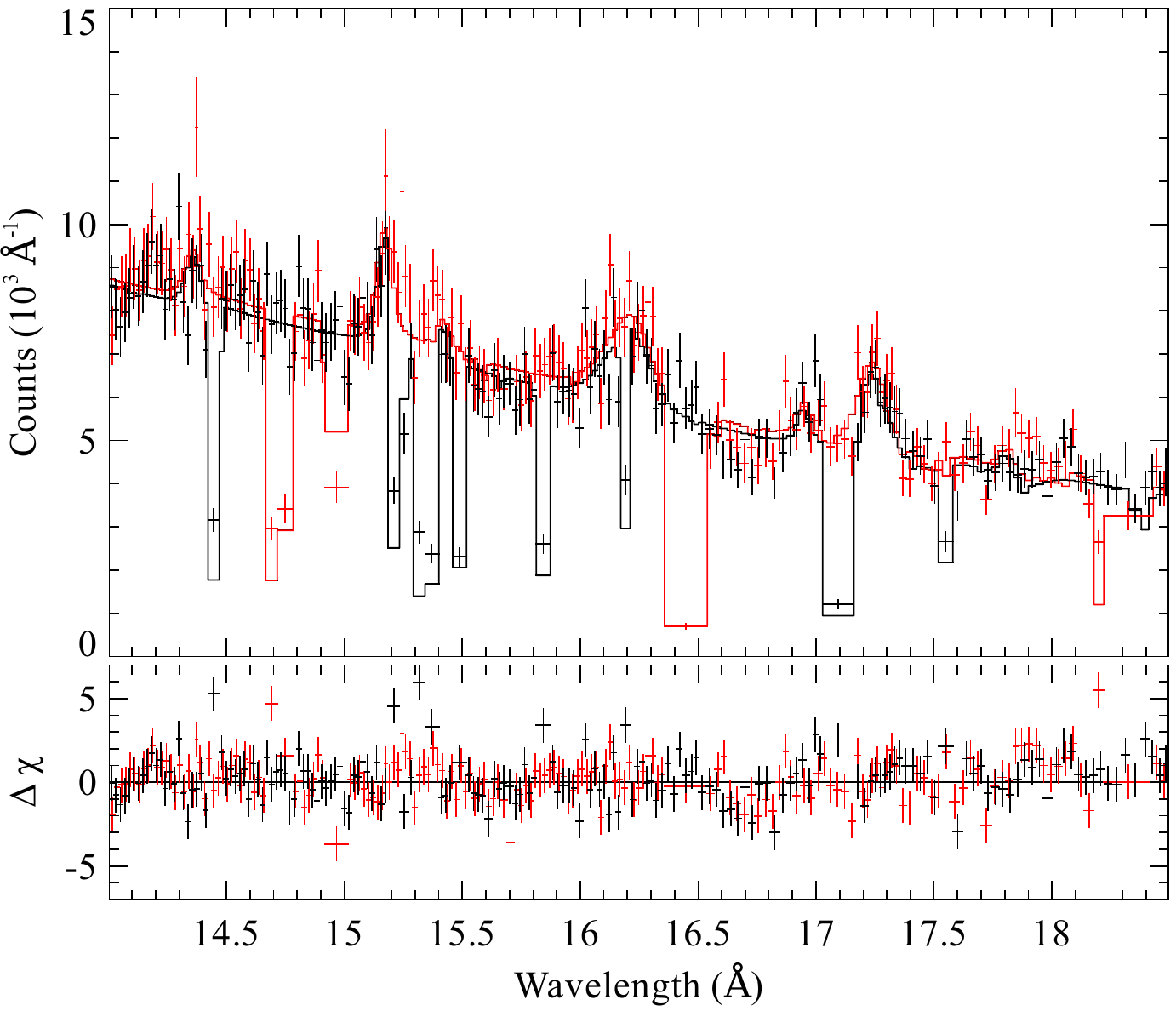}
  \caption{Absorbed powerlaw plus Gaussians model fit to the
    Fe~\textsc{xvii} region of the spectrum. The top panel shows the
    background-subtracted data and best fitting model. The bottom
    panel shows the contribution, $\Delta \chi$, of each channel to
    $\chi^2$. The RGS1 and RGS2 data and models are shown in black and
    red, respectively. The spectrum has been rebinned to a
    signal-to-noise ratio of 10 in \textsc{xspec}.}
  \label{fig:xvii_spec}
\end{figure}

\subsubsection{Fe \textsc{xvii}}
To examine the Fe~\textsc{xvii} line powers we fitted a spectral model
made up of an absorbed powerlaw plus Gaussians, fitted to the first
order spectra between 14 and 18.5{\AA} extracted from 99~per~cent of
the PSF. We used a Gaussian for each of the distinct Fe~\textsc{xvii}
lines (Table \ref{tab:mdotlines}), two Gaussians at 16.07 and
16.00{\AA} for Fe~\textsc{xviii} and O~\textsc{viii}, and
Fe~\textsc{xviii} Gaussians at 14.21{\AA} and 17.62{\AA}. The strong
lines at 15.01, 16.00, 16.07, and 17.06 were allowed to have variable
widths. The other lines were fixed at zero width. Galactic absorption
was fixed at $8.56 \times 10^{20}\psqcm$.

Table \ref{tab:mdotlines} shows the best fitting widths and
de-absorbed line luminosities. Fig.~\ref{fig:xvii_spec} shows the data
with best fitting model. The line powers would be increased if there
were any internal absorption within the cluster (as suggested by
\citealt{Crawford05} and \citealt{SandersReson06}). We convert these
powers into mass deposition rate by comparing them against the
expected flux from a cooling flow model. This was done by firstly
simulating a high signal-to-noise spectrum using a $10\Msunpyr$
\textsc{vmcflow} model, cooling between 4.5 and 0.0808~keV, with Solar
abundance, absorbed with Galactic absorption. We then measured the
fluxes in the simulated lines by fitting the same absorbed powerlaw
plus Gaussian model as we did to the data. The ratio of the line flux
in the data compared to the simultated model was used to obtain a mass
deposition rate.  Note that the assumed relative abundance can make a
substantial difference to these values. The mass deposition rates
obtained for three of the lines agree at $\sim 1.6 \Msunpyr$, but the
17.06{\AA} blended line is substantially stronger than the others.

\begin{figure}
  \includegraphics[width=\columnwidth]{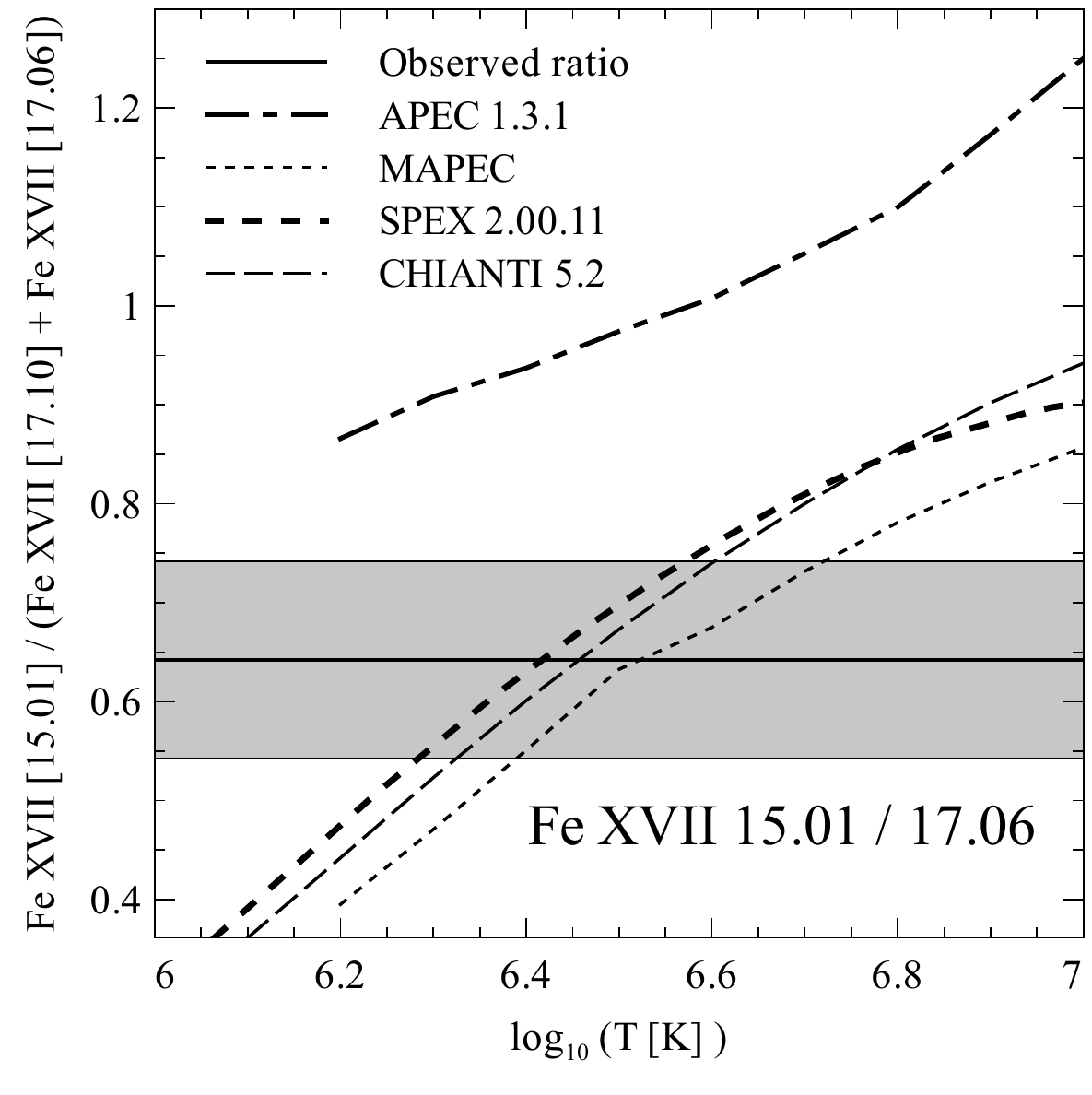}
  \caption{Predicted ratio of the 15.01 line to the blended 17.06 line
    as a function of temperature for a variety of different plasma
    codes. The line with surrounding shaded region shows the best
    fitting value and its uncertainty.}
  \label{fig:xvii_ratio}
\end{figure}

The \textsc{apec} model predicts that the 15.01{\AA} line should be
more powerful than the 17.06{\AA} blended line over a wide range of
temperatures.  However in our data the 17.06{\AA} lines are 56
per~cent brighter than the 15.01{\AA} line. A possible cause for this
is resonant scattering, as the 15.01{\AA} transition is resonant with
a large oscillator strength. If this were the case the scattered
radiation would have to be absorbed, as scattering only increases the
spectral width of the line, redistributing the flux.

However, a likely reason for the discrepancy in brightness may be due
to uncertainties in spectral models. We plot in
Fig.~\ref{fig:xvii_ratio} the predicted ratio of the 15.0 to blended
17.0{\AA} lines versus the temperature for some different plasma
codes, including \textsc{apec}, \textsc{mapec} (modified \textsc{apec}
results using a many-body perturbation theory method to calculate new
values for the emissivities of Fe and Ni L-shell lines;
\citealt{Gu07}), \textsc{spex} (the latest \textsc{mekal} results from
the \textsc{spex} package; \citealt{Kaastra00}) and \textsc{chianti}
\citep{Dere97,Landi06}.

It appears \textsc{apec} cannot reproduce the observed line ratio (as
found by \citealt{Xu02}). The other plasma codes are able to and agree
fairly well. The line ratio appears to indicate that the
Fe~\textsc{xvii} line emission comes from an average temperature of
$0.25^{+0.20}_{-0.09} \keV$. Note that although \textsc{apec} appears
to fit the data worse here, its quality of fit to the total spectra is
at least $\Delta \chi^2 \sim 300$ better than the \textsc{spex}
\textsc{mekal} model overall.

The Fe~\textsc{xvii} 17.06{\AA} line width of 0.047{\AA} translates
into a HEW of 36~arcsec (including the effect of the mirrors). This is
larger than the width of the emission line in the cross-dispersion
direction (Fig.~\ref{fig:lineprof_fe}; compatible with the PSF). The
longer observation analysed here has the cross-dispersion direction
placed along the plume. The difference in source extent between the
line widths and cross-dispersion profiles may be due to the very
coolest gas being extended in the direction perpendicular to the
cross-dispersion direction (Fig.~\ref{fig:sizemaps};
\citealt{Crawford05}).

If a cooling flow model plus powerlaw continuum is fit to this region
of the spectra containing the Fe~\textsc{xvii} lines, the best fitting
mass deposition rate is $1.90 \pm 0.13 \Msunpyr$. This is for a
cooling flow at Solar abundance cooling from 4.5 to 0.0808~keV. As the
Fe metallicity decreases, the mass deposition rate climbs to $3.8
\Msunpyr$.

\subsubsection{O \textsc{vii}}
We next placed upper limits on the O~\textsc{vii} lines at 21.6 and
22.1{\AA}. We fitted an absorbed powerlaw to the spectra from 19.5 to
23{\AA} plus two zero-width Gaussians at the line positions. We
calculated upper limits for the line powers and converted these to
upper limits on the mass deposition rate. These upper limits of $\sim
3 \Msunpyr$ are compatible with the Fe~\textsc{xvii} results, although
Solar metallicity is assumed. The O/Fe metallicity will affect these
values as Fe is the main coolant in this temperature range
\citep{BohringerHensler89}.

\begin{figure}
  \includegraphics[width=\columnwidth]{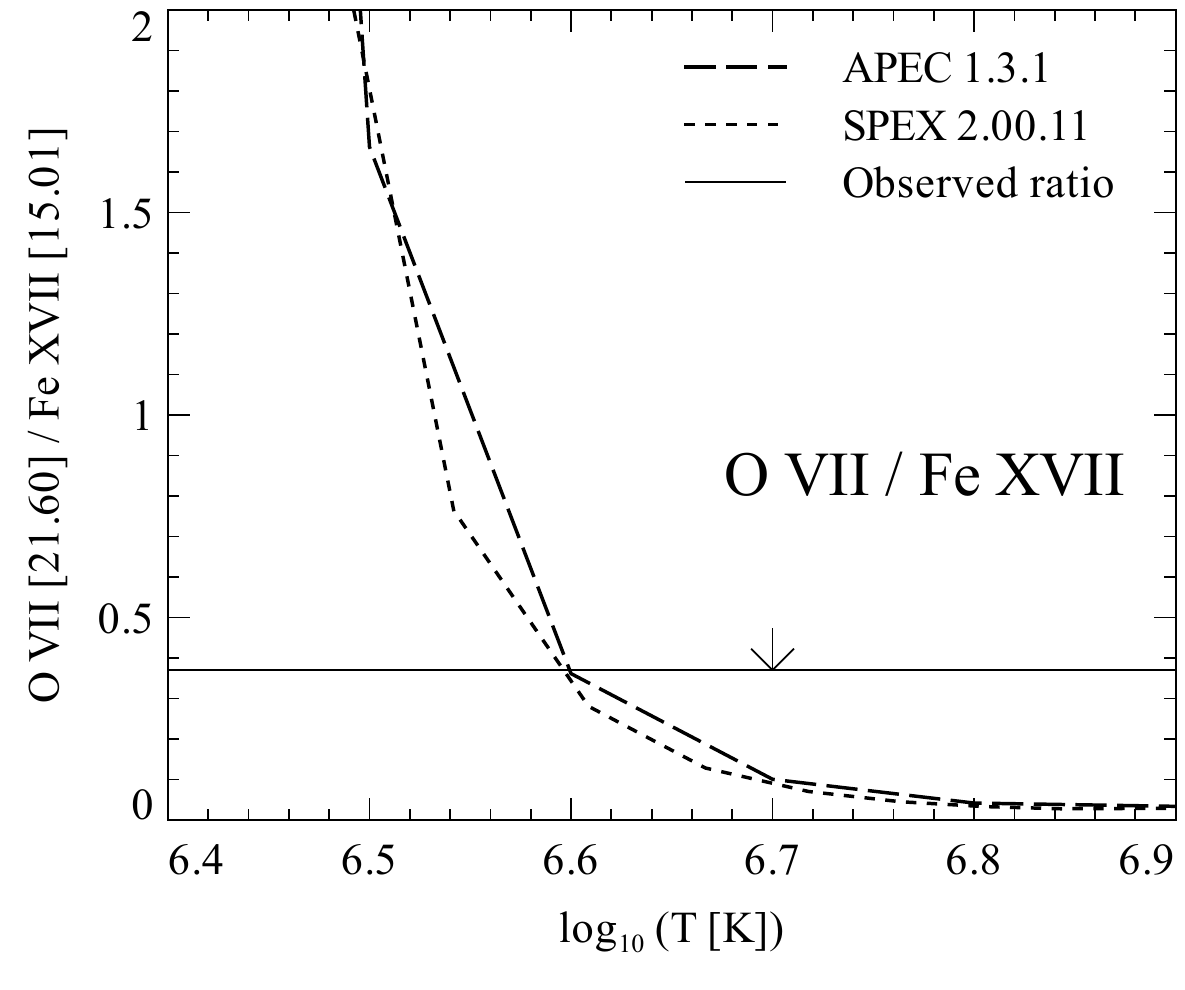}
  \caption{Flux ratio of 21.60{\AA} O~\textsc{vii} ratio to
    15.01{\AA} Fe~\textsc{xvii} line. The model lines show the ratio
    as function of temperature for Solar abundance. The horizontal
    line shows the observed upper limit.}
  \label{fig:ovii_ratio}
\end{figure}

We note that if the emission only came from gas at 0.25~keV, we would
expect the O~\textsc{vii} lines to be much stronger than
Fe~\textsc{xvii} (see Fig.~\ref{fig:spectrumzoom} lower panel). The
relative strength of the Fe~\textsc{xvii} and O~\textsc{vii} will
depend on the metallicity and ionisation balance, however. We show in
Fig.~\ref{fig:ovii_ratio} the expected ratio of the 21.60{\AA}
O~\textsc{vii} to 15.01{\AA} Fe~\textsc{xvii} line as a function of
temperature assuming Solar abundance. The ratio of any of the
O~\textsc{vii} to Fe~\textsc{xvii} lines is a strong function of
temperature. For sensible ranges in abundance (O/Fe from 0.3 to 1
Solar), it appears unlikely that the temperature of the material
emitting the Fe~\textsc{xvii} lines is less than $10^{6.55} -
10^{6.6}$~K. O/Fe of around 0.1 Solar is required for 0.25~keV
material to be consistent with the line ratio. This means a likely
lower temperature is $0.3-0.35\keV$, still within the allowable range
of the Fe~\textsc{xvii} ratio. These estimates all assume that the
line emission comes from isothermal gas.

\begin{figure}
  \includegraphics[width=\columnwidth]{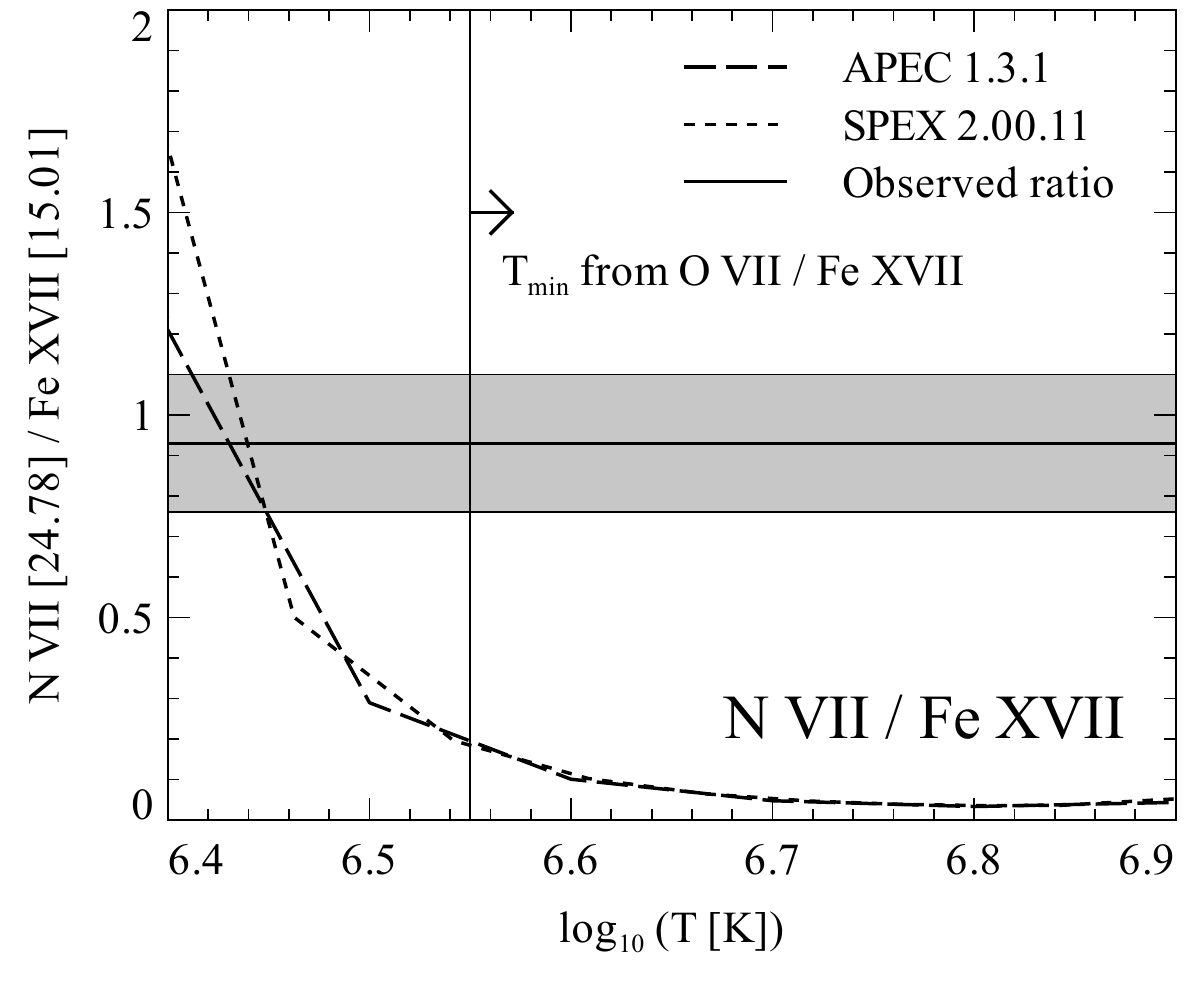}
  \caption{Flux ratio of 24.78{\AA} N~\textsc{vii} ratio to 15.01{\AA}
    Fe~\textsc{xvii} line. The model lines show the ratio as function
    of temperature for Solar abundance. The shaded region shows the
    observed ratio. The vertical line shows the minimum temperature
    from the O~\textsc{vii} to Fe~\textsc{xvii} line ratios
    (Fig.~\ref{fig:ovii_ratio}).}
  \label{fig:nvii_ratio}
\end{figure}

We also note that resonant scattering could be important for the
21.6{\AA} O\textsc{vii} line, which has a high oscillator strength
(0.7). Although scattering by itself would not decrease the line flux,
it would broaden the spectral line, making it harder to
detect. Absorbing material within the scattering region could
also reduce the line flux.

\subsubsection{N \textsc{vii}}
We also try to measure the mass deposition rate in the absence of
heating using the N~\textsc{vii} line. We fit the region 23 to 30{\AA}
with an absorbed powerlaw plus Gaussian with variable width. The power
in this line compared to the Fe~\textsc{xvii} lines is large: assuming
solar abundance its power corresponds to a $10\Msunpyr$ cooling
flow. This indicates that the N abundance of the cool X-ray emitting
gas relative to Fe is large, as was also shown by the spectral
fits. Additionally the N~\textsc{vii} line covers a larger range in
temperature, and so is sensitive to hotter gas which shows a higher
effective mass deposition rate (Fig.~\ref{fig:mdott}). We note that
any additional absorption will increase this line power relative to
the Fe~\textsc{xvii} lines.

The spectral width of the line is also very close to the
Fe~\textsc{xvii} 17.06{\AA} line width, meaning it originates from the
same spatial region. The finite widths of the lines (a point source
would provide an upper limit as the response already takes account of
the \emph{XMM} PSF) show that the emitting region is resolved, with a
HEW of around 35~arcsec (taking into account the \emph{XMM} PSF HEW).

If the lower bound on the temperature is $10^{6.55}$~K, then the ratio
of the N~\textsc{xvii} to Fe~\textsc{xvii} 15.01{\AA} line implies
supersolar N abundances (Fig.~\ref{fig:nvii_ratio}). The minimum N/Fe
metallicity as indicated by the \textsc{apec} and \textsc{spex} model
line ratios is at least $4 \Zsun$. This value is in rough agreement
with the spectral fitting results.

\section{Discussion}

\subsection{Temperature range}
The observed range in temperature of X-ray emitting gas, from 3.7~keV
\citep{SandersCent02} to $\sim 0.35 \keV$ is one of the largest
observed up to now in a cluster of galaxies.  These RGS results
confirm earlier CCD detections of cool gas.

Another cluster with a large apparent X-ray emitting temperature range
is the Perseus cluster, containing X-ray emitting filaments at around
0.7~keV in a cluster up to 7~keV \citep{SandersPer07}, although these
are clearly influenced by magnetic fields.

The result emphasises that gas certainly does exist at temperatures
below the often quoted factor of 2 or 3 in galaxy
clusters. Fig.~\ref{fig:norms} shows that the emission measure of
cooler components is progressively smaller than a cooling flow model
as the temperature decreases. The lack of detections of cool gas in
other clusters using RGS observations could be due to their relatively
short exposures, which are typically only 30--40~ks
\citep{Peterson03}.

The large range in temperature in this cluster may be connected to the
lack of disruption in its core. The high enrichment of the central
parts of the cluster indicates it has been stable for 8~Gyr or more
\citep{SandersEnrich06}.

\subsection{Cooling flow model}
\begin{figure}
  \includegraphics[width=\columnwidth]{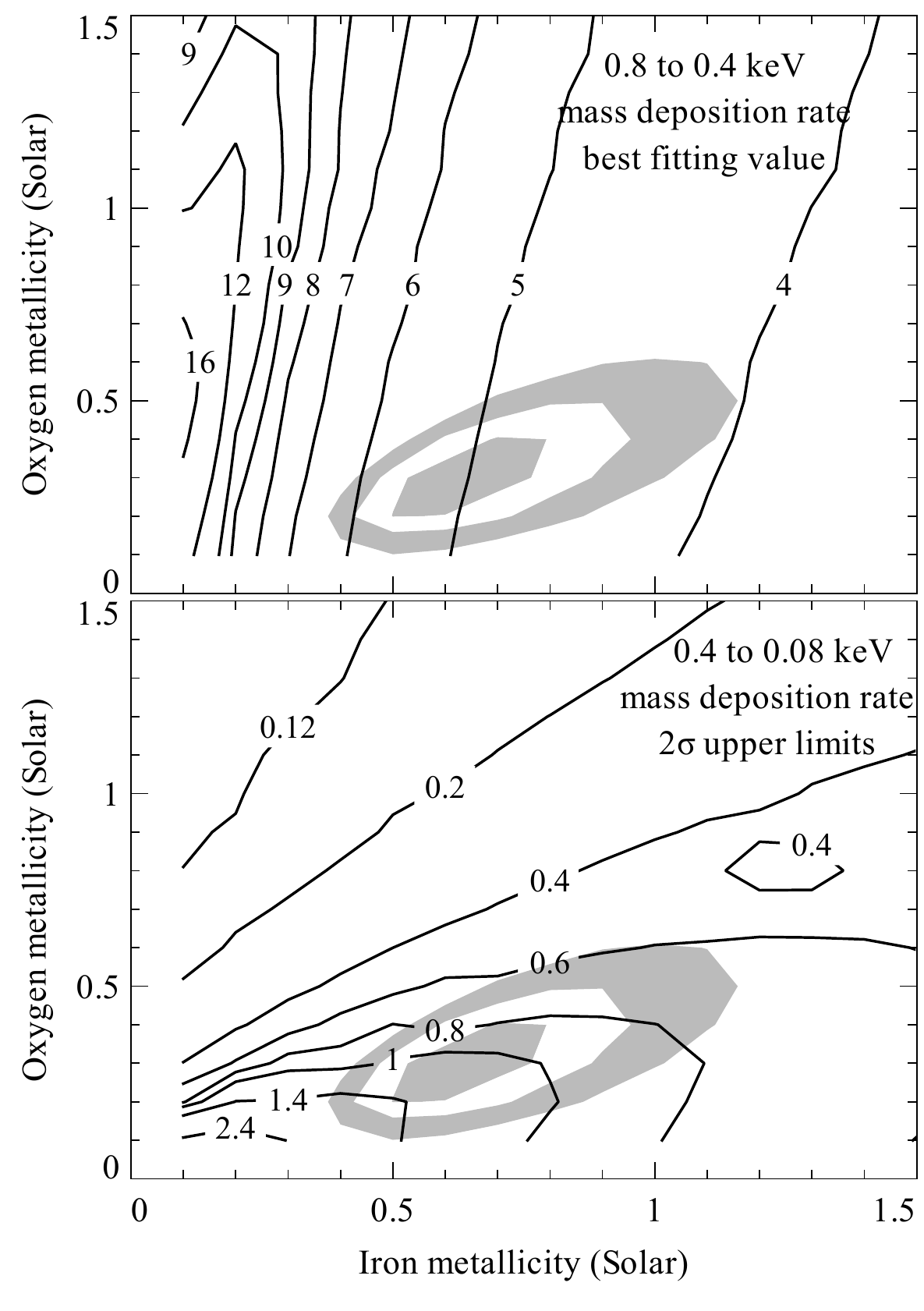}
  \caption{Variation of mass deposition rate with Iron and Oxygen
    metallicity for the two coolest temperature ranges in
    Fig.~\ref{fig:mdott}. The top panel shows as contour lines the
    best fitting mass deposition rates in \Msunpyr for the 0.8 to 0.4
    keV temperature range. The bottom panel shows upper limits for 0.4
    to 0.08 keV. Also plotted as shaded regions on both panels are the
    1, 2 and 3-$\sigma$ confidence regions for the O and Fe
    metallicity, calculated using the $\Delta \chi^2$ method from the
    minimum $\chi^2$ value.}
  \label{fig:mdotcont}
\end{figure}

\begin{figure}
  \includegraphics[width=\columnwidth]{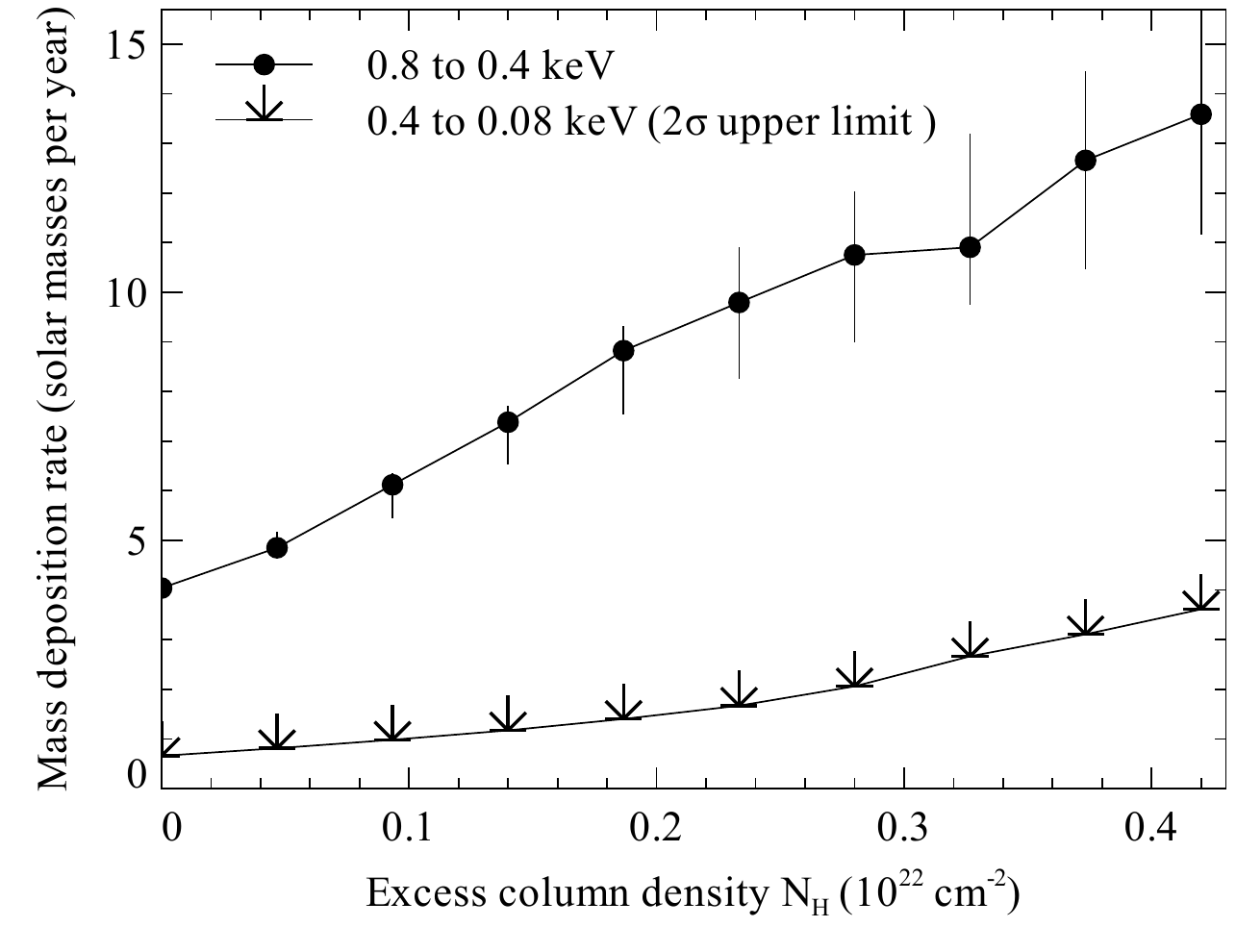}
  \caption{Variation of mass deposition rate with excess column
    density above $8 \times 10^{20} \psqcm$. Shown are the best
    fitting mass deposition rates in \Msunpyr for the 0.8 to 0.4 keV
    temperature range, or upper limits for the 0.4 to 0.08 keV
    temperature range.}
  \label{fig:nh_mdot}
\end{figure}

Fig.~\ref{fig:norms} and Fig.~\ref{fig:mdott} show that a full simple
cooling flow, where gas is cooling by radiative cooling at a constant
rate, is not operating in this cluster.  However there are some
possible reasons why the amount of cold gas may be underestimated:

\begin{enumerate}
\item These models assume that there is a fixed metallicity of
  material as a function of temperature. The O and Fe abundances
  change the cooling function and strengths of lines significantly.
  Fig.~\ref{fig:mdotcont} shows how the mass deposition rate for the
  $5 \times$\textsc{vmcflow} cooling flow model varies as a function
  of metallicity, for the two coolest temperature ranges. This was
  made by iterating over a grid of O and Fe abundances and measuring
  the mass deposition rate at each point. It shows that the real
  cooling rate is a strong function of metallicity. If the metallicity
  of the material in the central regions is small there could be
  significantly more cooling taking place. There is evidence for a
  drop in metallicity in the central regions
  \citep{SandersCent02}. This analysis highlights that the best
  fitting Fe and O metallicities for the two coolest components are
  small.

  Indeed, O~\textsc{vii} emission is the main indicator of cool gas in
  this waveband, but Fe line emission is actually the main coolant
  \citep{BohringerHensler89}. Therefore the O lines are suppressed if
  the Fe metallicity is increased while the O metallicity is constant,
  for a fixed observed Fe~\textsc{xvii} strength in a cooling
  flow. Therefore gas below 0.35~keV can exist without strong
  O~\textsc{vii} emission.

\item Intrinsic absorption could hide much of the cooling taking
  place. There is evidence that there are large amounts of obscuring
  material in the core of this cluster
  \citep{Crawford05,SandersReson06}, where the coolest material
  resides. It is difficult for spectral fitting to account for this as
  the continuum is hard to detect for low-temperature gas. We
  investigated this by fitting the $5\times$\textsc{vmcflow} to the
  data with varying levels of absorption applied to the two lowest
  temperature components (note that these results were obtained by
  varying the absorption separately for each component). In
  Fig.~\ref{fig:nh_mdot} we plot the increase in mass deposition rate
  it is possible to obtain by increasing the absorption.

\item The cooling flow model fitted to the data assumes that isobaric
  (constant pressure) cooling is taking place. If the cooling time
  becomes shorter than the sound-crossing time, then isochoric
  (constant density) cooling may be more appropriate. This could also
  result from the magnetic pressure in clumps becoming dominant as the
  cooling proceeds. The strength of UV emission lines from isochoric
  cooling is only around 60 per~cent of that expected from isobaric
  cooling \citep{EdgarChevalier86}. This would lead to the amount of
  cooling being underestimated if a cooling flow were taking place. As
  we show below, the mean radiative cooling time is only $\sim
  10^7$~yr for the very coolest X-ray detected gas.

\item If the X-ray emitting gas mixes with the cooler optical line
  emitting material in the very central regions \citep{Crawford05},
  than it will rapidly cool non-radiatively much faster than a cooling
  flow model would predict \citep{FabianCFlow02}. The thermal energy
  is eventually radiated at much longer wavelengths, in the
  optical/UV/IR filaments, for example, or by dust. There is, indeed,
  extended mid-IR emission observed from NGC~4696 \citep{Kaneda07},
  which could in part be powered by such a mixing process.
\end{enumerate}

When fitting the data with a cooling flow model, we find progressively
smaller mass deposition rates at lower temperatures, as found
previously by \cite{Peterson03}. However we detect gas over a much
wider range in temperature (a factor of 10 rather than 3) than in
their shorter observations, with a wider range of best fitting mass
deposition rates (a factor of 50 rather than 10), in reality
corresponding to detecting a wider range of emission measures for the
temperature components.

Taking gas at 0.35 keV in pressure equilibrium with the gas with
$n_\mathrm{e} \sim 0.1 \pcmcu$ and $kT \sim 1$~keV
\citep{SandersReson06}, implies a density of this cool material of
around $n_\mathrm{e} \sim 0.3 \pcmcu$. We can calculate the bolometric
luminosity from a unit volume of this material (using the abundances
from the $5\times$\textsc{vapec} model) and its internal energy
(assuming $\frac{5}{2} \, k_\mathrm{B}T n_\mathrm{tot}$). Dividing the
two gives a mean radiative cooling time of only $\sim 10^7\yr$ for
this 0.35~keV material.

The 0.4~keV component in the $5\times$\textsc{vapec} model is close to
0.35~keV in temperature. Assuming the material is in pressure
equilibrium with its surroundings and taking its emission measure from
the $5\times$\textsc{vapec} spectral fit, we can estimate it occupies
a volume of $2.3\kpc^3$. The region that the Fe~\textsc{xvii} lines
are emitted from has a radius of at most 10~arcsec, implying a maximum
volume of $120\kpc^3$. This means the 0.4~keV material must fill only
2~percent of that very central region. \emph{Chandra} temperature maps
of the core show that the coolest gas is not volume-filling, but
exists in the form of cool blobs \citep{Crawford05}.

\subsection{Heating}
\label{sect:heating}

\begin{figure}
  \includegraphics[width=\columnwidth]{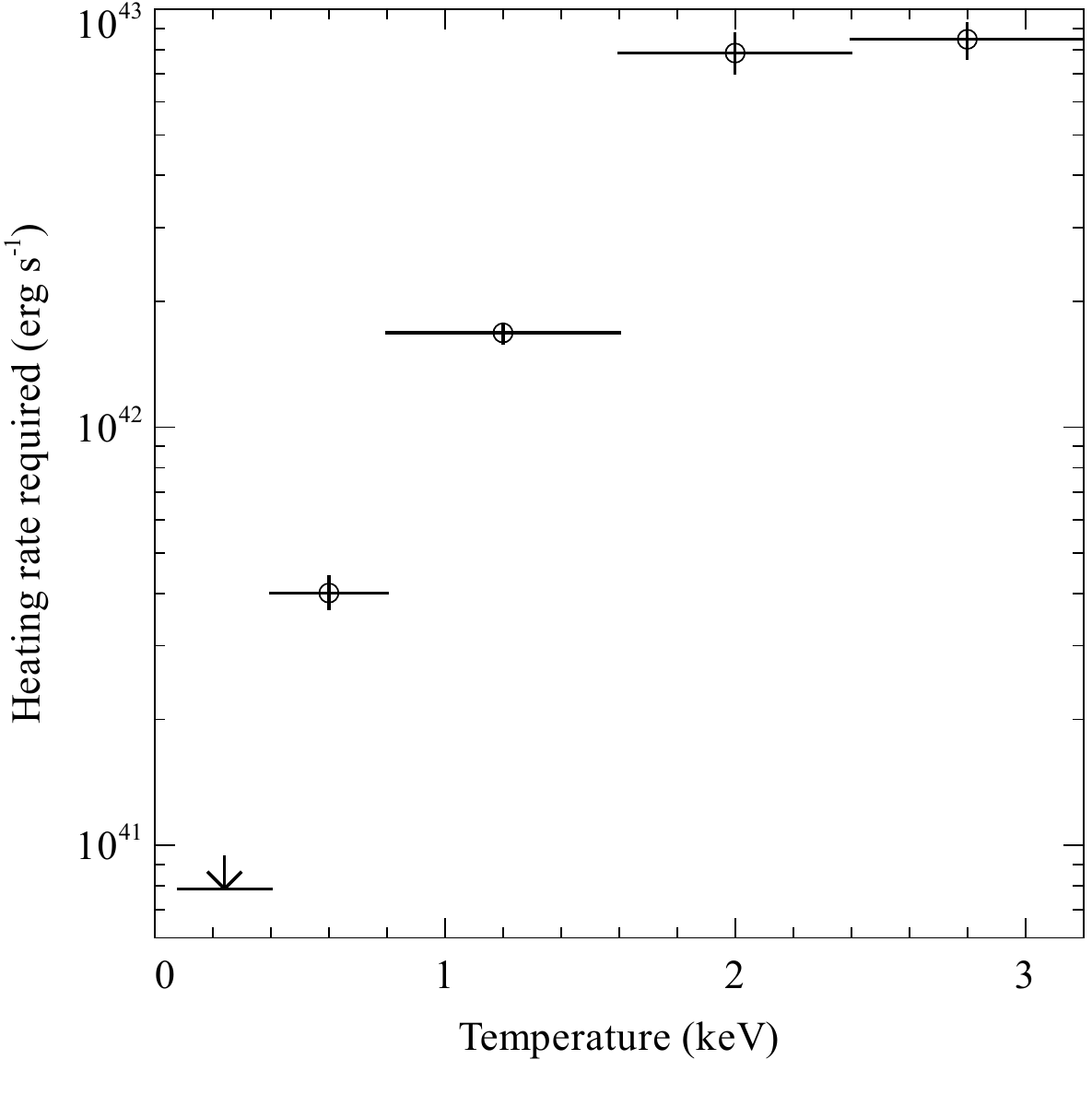}
  \caption{Heating rate required to combat cooling, calculated for
    each of the individual temperature ranges in the
    $5\times$\textsc{vmcflow} model (see Fig.~\ref{fig:mdott}).}
  \label{fig:mdotheating}
\end{figure}

The amount of heating which would be required to offset cooling of the
ICM in the cluster can be estimated. We took the
$5\times$\textsc{vmcflow} cooling flow model fits to the spectra and
generated a chain of parameters using a Markov Chain Monte Carlo. The
chain was created with the built-in \textsc{xspec} functionality and
was checked for convergence by repeating the analysis with a different
starting point. We generated luminosities for each of the sets of
parameters in the chain for the individual temperature ranges in the
model. We examined the distribution of luminosities to estimate the
medians and uncertainties. The luminosities of each temperature range
are the rates of heating required to offset cooling.

In Fig.~\ref{fig:mdotheating} we show the amount of heating required
for each temperature range of gas in order to combat cooling. At
temperatures of above $\sim 1.5 \keV$, significant heating rates of a
few times $10^{43} \ergps$ are required. For the coolest gas it is
likely that mixing and conduction complicate this picture. The cool
gas we observe here must exist as distributed clumps and the short
radiative cooling time places constraints on heating models.

\subsection{Metallicity of the gas}
\label{sect:metals}
The best fitting metallicities we measure show some model
dependence. Those models with larger number of temperature components
tend to have higher metallicites.  However there is quite good
agreement between the ratios of the various elements between the best
fitting models here and with the previous CCD results.

The metallicities we measure here are smaller than the peak values
obtained using CCD data from \emph{Chandra} and \emph{XMM}
\citep{SandersEnrich06}. It is however difficult to compare them
exactly as the CCD results are spatially-resolved and the metallicity
appears to rise to high values, but drops in the very central
region. The temperature is also declining towards the centre. Over the
region examined by the RGS detectors, it is likely that there is a
rise in metallicity with radius.

Ni shows significantly smaller metallicities of around $2.4\Zsun$ here
when compared to earlier \emph{Chandra} and \emph{XMM}
spatially-resolved CCD results (which peak near $4 \Zsun$;
\citealt{SandersEnrich06}). This can be mostly resolved by allowing
the Ni to vary between the coolest and hotter components, obtaining
values of $3.5^{+0.5}_{-0.4}$ and $1.7_{-0.1}^{+0.2} \Zsun$,
respectively (for the $5 \times$\textsc{vapec} model). This highlights
that the assumptions or choices we made in modelling the data can have
a large impact on the obtained quantities.

The generally smaller metallicites we find here, when compared to CCD
results, may be due to problems modelling the continuum from hot gas
outside of the central region. Fig.~\ref{fig:norms} shows that the RGS
results require very little hot gas in the spectral fits when compared
to \emph{Chandra}. If we add the expected 3--4~keV hot component from
the \emph{Chandra} spectral fits to the RGS spectral models, freezing
its emission measure to the same value, then we typically find the
metallicities we measure are increased by 30 to 40~percent. The RGS
spectra appear to require much less of this hot material than all
previous analyses. Therefore there may be calibration uncertainties at
short wavelengths in the RGS spectra.

Another possible reason for higher continuum, leading to
underestimates of the metallicity, would be the previously observed
non-thermal emission \citep{Allen00,DiMatteo00}. The sedimentation of
helium in the core of the cluster would also lead to higher continuum
\citep{EttoriFabian06}.

\begin{figure}
  \includegraphics[width=\columnwidth]{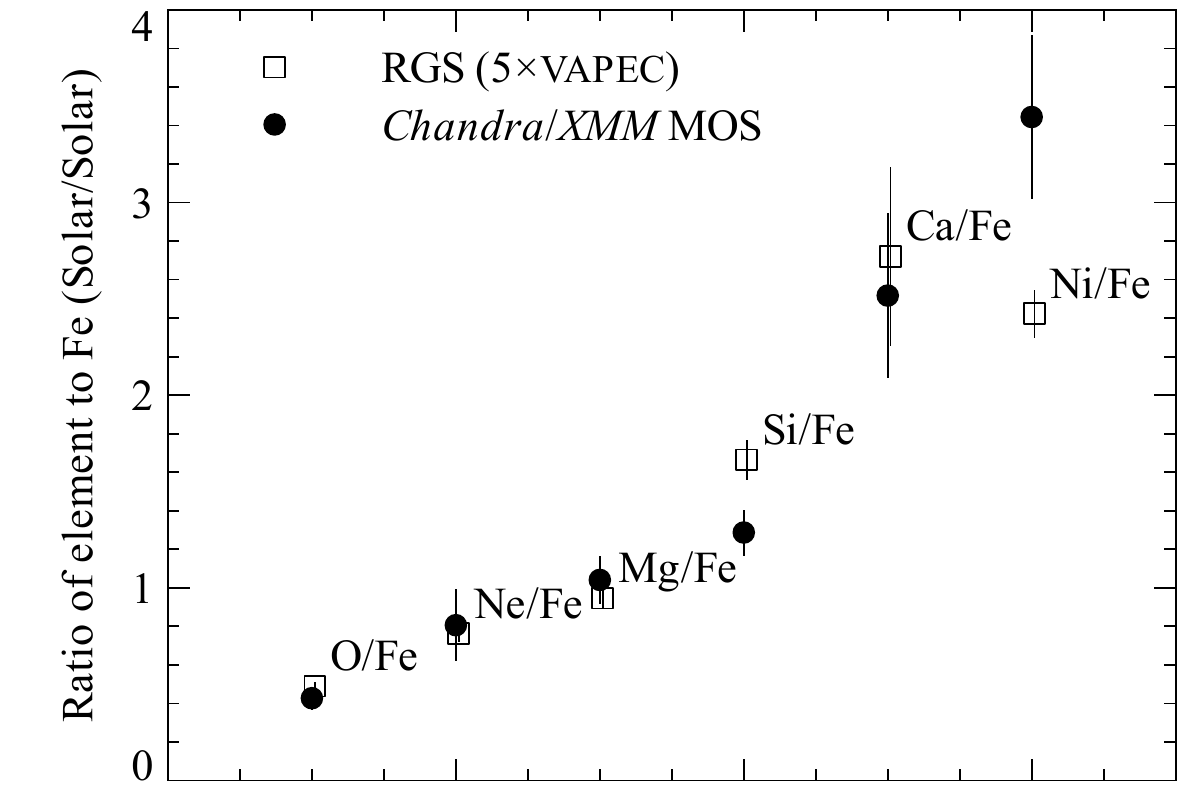}
  \caption{Ratio of metallicity of elements to Fe obtained from the
    RGS data with the $5\times$\textsc{vapec} model, compared to the
    results from the central regions of the cluster from combined
    \emph{Chandra} and \emph{XMM} EPIC MOS data
    \citep{SandersEnrich06}.}
  \label{fig:metalratios}
\end{figure}

The ratio of elemental abundances is likely to be more robust than
measurements of individual abundances. In Fig.~\ref{fig:metalratios}
we plot the ratio of the metallicity of various elements to Fe (the
ratios are $\Zsun/\Zsun$), measured using the multitemperature
$5\times$\textsc{vapec} model (Fig.~\ref{fig:metals}). Also shown are
the same ratios obtained from \emph{Chandra} and \emph{XMM} EPIC MOS
data. These were calculated by averaging the central \emph{XMM} and
\emph{Chandra} East and West points from figure~15 (bottom panel) in
\cite{SandersEnrich06}. The plot shows that we get good agreement
between the RGS, and previous CCD ratios within the inner 10~kpc
radius, with some discrepancy for Si/Fe and Ni/Fe. These results
indicate that the CCD-derived metallicity ratios in
\cite{SandersEnrich06} are robust.

If the coolest gas is examined separately from the hotter gas, it
appears to have lower Fe and O metallicity, confirming the central
abundance drop in the spatially resolved CCD
measurements. Fig.~\ref{fig:mdotcont} shows the best-fitting values of
the cooler gas are only around 60~per~cent of the metallicity of the
hotter gas.

It is possible that the metallicity variations are with temperature,
not radius, although we do not understand why. If there is a continued
decrease of metallicity as temperature decreases then the low
O~\textsc{vii} result might be understood. It is unlikely that the low
metallicity gas can be introduced into the centre, and we presume that
the metals are removed. At a temperature of 0.2~keV the cooling rate
is sensitive to the metallicity. One simple explanation is that the
high metallicity gas is inhomogeneous, conduction is suppressed
between clumps, leading to the the high metallicity regions cooling
out \citep{MorrisFabian03}. This means that both the injection of
metals is a inhomogeneous process and the metals remain poorly mixed
for a considerable time (several Gyr).

We measure values for the N abundance of between 1.6 (from the
$5\times$\textsc{vapec} model) to $4\Zsun$ (for the
$5\times$\textsc{vmcflow} and lower limit from the line ratios).

The nitrogen emission lines in the optical spectrum of the
emission-line filamentary nebula in NGC\,4696, the central galaxy in
the Centaurus cluster also indicate a high nitrogen
abundance. Specifically, the [NII]/H$\alpha$ line ratio found by
\cite{Johnstone87} of 3.2 is higher than the ratio of less than two
which can be accounted for by photoionization or shock models for a
solar abundance plasma
\citep{Heckman89,VoitDonahue90,CrawfordFabian92}. A high nitrogen
abundance in both the optical filaments and the hot gas supports the
hypothesis that they have a common origin, with the filaments
representing gas which has cooled from the hotter phase.

\section{Conclusions}
We clearly detect Fe~\textsc{xvii} emission from the core of the
Centaurus cluster. Fitting spectral models and measuring emission line
ratios shows that the temperature of the gas declines to 0.3 to
0.45~keV. These results from this deep RGS observation show the widest
range in ICM temperature detected in a cluster of galaxies, with a
factor of 10 or more.

The data confirm that the metallicity of the ICM declines in the very
central regions. Whether this is a real decline in metallicity, the
result of an inhomogeneous metallicity distribution
\citep{MorrisFabian03} or excess continuum is unclear. We do, however,
find strong nitrogen enhancement in the inner 6~kpc radius.
 
The very coolest gas is concentrated in the centre of the cluster, as
previously found by \emph{Chandra}, but there is much less cool
material than would be expected from a simple cooling flow.

The problem of how to prevent this very cool gas, which has a mean
radiative cooling time of $10^7$~yr, from cooling remains.  Any
mechanism which heats the gas must preserve the sharp metallicity
gradients and not switch off for timescales of greater than
$10^7$~yr. Alternatively, the ICM in the cluster may be cooling in a
non-radiative way below 0.4~keV. If this material formed stars, the
rate of mass deposition is compatible with the metallicity of the ICM
\citep{SandersEnrich06}.

Using this very deep observation we are able to measure the amount of
cool X-ray emitting gas very precisely in a cluster core. In order to
get a more complete picture, it is necessary to make deep observations
of several nearby clusters to see whether the large ranges in
temperature are found in all objects.

\section*{Acknowledgements}
ACF acknowledges the Royal Society for support. SWA and RGM
acknowledge support by the U.S. Department of Energy under contract
number DE-AC02-76SF00515, and by the National Aeronautics and Space
Administration through \emph{XMM-Newton} Award Number
NNX06AG29G. Based on observations obtained with \emph{XMM-Newton}, an
ESA science mission with instruments and contributions directly funded
by ESA Member States and NASA.

\bibliographystyle{mnras}
\bibliography{refs}

% required because of bug in MN2e style file
% throws away figs otherwise
\clearpage

\end{document}

%% file: defn.tex
\newcommand{\Mpc}{\rm\thinspace Mpc}
\newcommand{\kpc}{\rm\thinspace kpc}

\newcommand{\km}{\rm\thinspace km}

\newcommand{\cm}{\rm\thinspace cm}

\newcommand{\pcmcu}{\hbox{$\cm^{-3}\,$}}

\newcommand{\yr}{\rm\thinspace yr}

\newcommand{\s}{\rm\thinspace s}

\newcommand{\Msun}{\hbox{$\rm\thinspace M_{\odot}$}}

\newcommand{\Msunpyr}{\hbox{$\Msun\yr^{-1}\,$}}
\newcommand{\keV}{\rm\thinspace keV}

\newcommand{\erg}{\rm\thinspace erg}

\newcommand{\ergps}{\hbox{$\erg\s^{-1}\,$}}

\newcommand{\kmps}{\hbox{$\km\s^{-1}\,$}}

\newcommand{\kmpspMpc}{\hbox{$\kmps\Mpc^{-1}$}}
\newcommand{\Zsun}{\hbox{$\thinspace \mathrm{Z}_{\odot}$}}

\newcommand{\psqcm}{\hbox{$\cm^{-2}\,$}}

\newcommand{\ps}{\hbox{$\s^{-1}\,$}}